\documentclass[journal]{IEEEtran}
\ifCLASSINFOpdf
  \usepackage[pdftex]{graphicx}
  \graphicspath{../eps/}
  \DeclareGraphicsExtensions{.pdf,.jpeg,.png}
\else
  \usepackage[dvips]{graphicx}
  \graphicspath{{../eps/}}
  \DeclareGraphicsExtensions{.eps}
\fi
\usepackage{siunitx}
\usepackage[colorlinks=false]{hyperref}
\usepackage{color}
\usepackage{cite}
\usepackage{amsmath}
\usepackage{amssymb}
\usepackage{mathptmx}
\usepackage{epsfig}
\usepackage{times}
\usepackage{threeparttable}
\usepackage{stfloats}
\usepackage{amsmath,amssymb,amsfonts}
\usepackage{algorithmic}
\usepackage{textcomp}
\usepackage{array}

\usepackage{array}

  \usepackage[caption=false,font=footnotesize]{subfig}
\begin{document}
%
\title{MagNI: A Magnetoelectrically Powered and Controlled Wireless Neurostimulating Implant}
%
%
%

\author{
        Zhanghao Yu*,~\IEEEmembership{Student Member,~IEEE,}
        Joshua C. Chen*, 
        Fatima T. Alrashdan,~\IEEEmembership{Student Member,~IEEE,}
        Benjamin~W. Avants,
        Yan He,~\IEEEmembership{Student Member,~IEEE,}
        Amanda Singer, 
        Jacob T. Robinson,~\IEEEmembership{Senior Member,~IEEE,}
        and~Kaiyuan~Yang,~\IEEEmembership{Member,~IEEE}
\thanks{This work was supported in part by the National Institute of Biomedical Imaging and Bioengineering of the National Institutes of Health under award U18EB029353 and the National Science Foundation under award ECCS-2023849. (Corresponding Authors: Kaiyuan Yang, kyang@rice.edu; and Jacob~T.~Robinson, jtrobinson@rice.edu)}
\thanks{All authors are with Rice University, Houston, TX 77005, USA. J. T. Robinson is also with Baylor College of Medicine, Houston, TX 77030, USA.}
\thanks{(*Zhanghao Yu and Joshua C. Chen contributed equally to this work.)}
\thanks{\copyright 2020 IEEE. Personal use is permitted, but republication/redistribution requires IEEE permission.}
}
\markboth{IEEE TRANSACTIONS ON BIOMEDICAL CIRCUITS AND SYSTEMS}%
{Shell \MakeLowercase{\textit{et al.}}: Bare Demo of IEEEtran.cls for IEEE Journals}
%



\maketitle

\begin{abstract}
This paper presents the first wireless and programmable neural stimulator leveraging magnetoelectric (ME) effects for power and data transfer. 
Thanks to low tissue absorption, low misalignment sensitivity and high power transfer efficiency, the ME effect enables safe delivery of high power levels (a few milliwatts) at low resonant frequencies (\si{\sim}250~kHz) to mm-sized implants deep inside the body (30-mm depth). 
The presented MagNI (Magnetoelectric Neural Implant) consists of a 1.5-mm$^2$ 180-nm CMOS chip, an in-house built 4~×~2~mm ME film, an energy storage capacitor, and on-board electrodes on a flexible polyimide substrate with a total volume of 8.2~mm$^3$.
The chip with a power consumption of 23.7~\si{\micro}W includes robust system control and data recovery mechanisms under source amplitude variations (1-V variation tolerance). 
The system delivers fully-programmable bi-phasic current-controlled stimulation with patterns covering 0.05-to-1.5-mA amplitude, 64-to-512-\si{\micro}s pulse width and 0-to-200-Hz repetition frequency for neurostimulation. 
\end{abstract}


%
\IEEEpeerreviewmaketitle

\section{Introduction}
%
%
%
%
\IEEEPARstart{N}{eurostimulation} holds significant promise as a tool to modulate nerves for both neuroscience research and clinical therapies. Peripheral nerve stimulation (PNS) is a common approach to treat neuropathic pain. For example, devices can be implanted to deliver electrical pulses to the spinal cord, which can help prevent pain signals from reaching to the brain \cite{shealy_electrical_1967, verrills_review_2016}. 

\begin{figure}[t]
      \centering
      \includegraphics[width=0.8\linewidth]{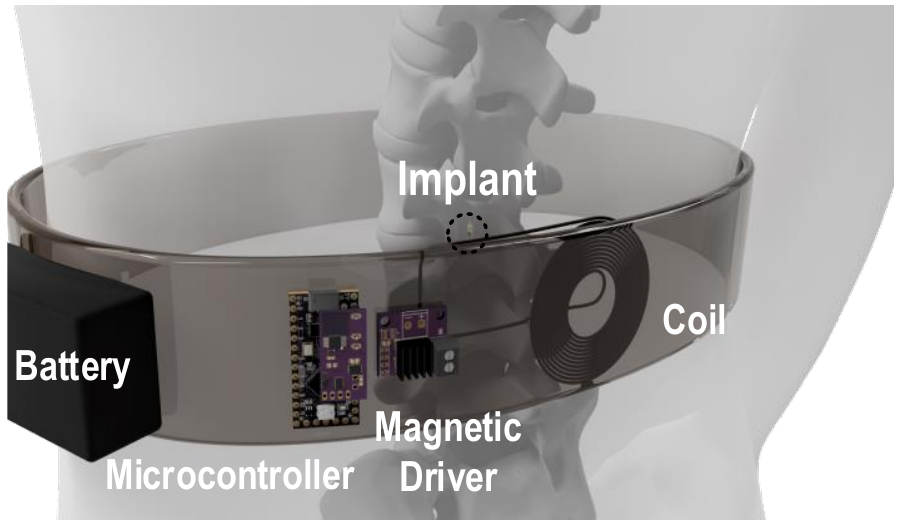}
      \caption{Conceptual diagram showing a wearable spinal cord neurostimulation system for pain relief, the implant is remotely powered via magnetic fields.}
      \label{S1_Concept}
   \end{figure}

A fundamental challenge in developing miniature neural implants is delivering power to devices inside the body. The use of a wired power supply causes failures for neural implants, as lead wires increase the risks of infections, restrict device deployment and affect subject mobility \cite{hargreaves_kirschner_2004, biran_brain_2007}. Batteries add considerable weight and increases device footprint \cite{pinnell_miniaturized_2018}. They are also required to be replaced or recharged frequently, which limits their long-term clinical applications. Compared to wired or battery powered implants, wirelessly powered battery-free neural stimulators have the potential to provide less invasive, longer lasting interfaces to nerves. Ideally, these implants would be programmable so that they can be reconfigured to suit user needs. Fig.~\ref{S1_Concept} illustrates the concept of the proposed spinal cord stimulation system. The wireless implant receives power and data from the portable battery powered transmitter (TX) via magnetic field; the microcontroller, magnetic field driver, and the battery are assembled in a wearable belt. 

    \begin{figure}[t]
      \centering
      \includegraphics[width=0.8\linewidth]{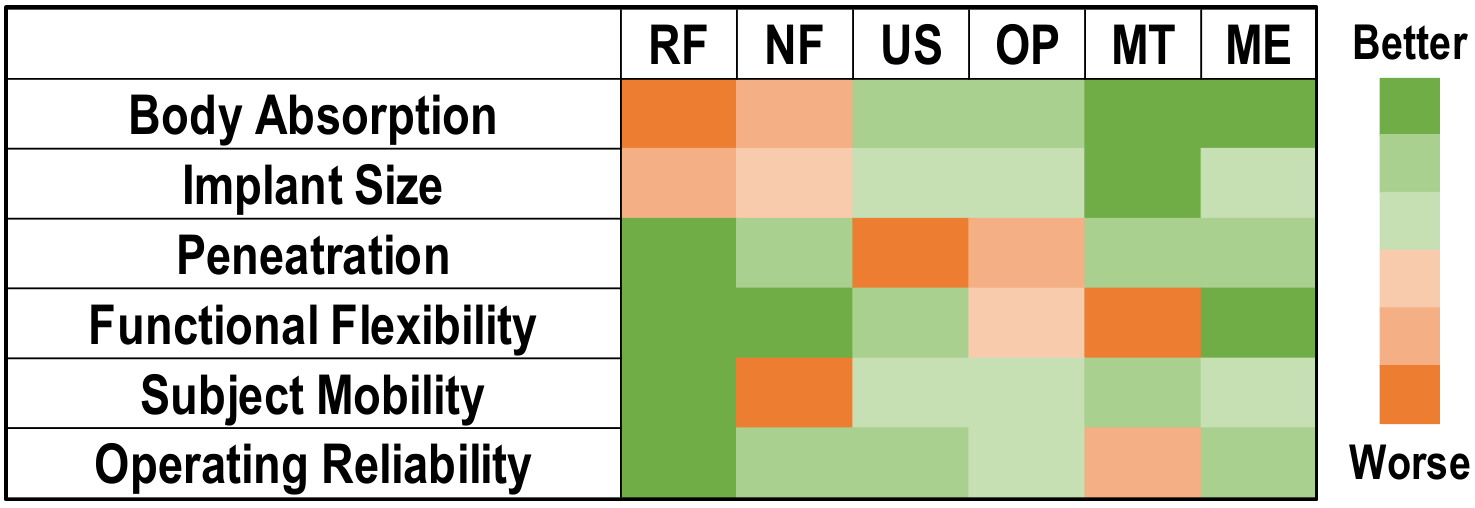}
      \caption{Comparison of wireless power transfer modalities for bioelectronic implants. (RF: radio-frequency field. NF: near-field inductive coupling. US: ultrasound. OP: optoelectronics. MT: magnetothermal nanoparticles. ME: magnetoelectric effects.)}
      \label{S1_Comp}
   \end{figure}

While various wireless neural implants exploiting radio-frequency (RF) electromagnetic (EM)~\cite{montgomery_wirelessly_2015, leung_cmos_2018}, inductive coupling~\cite{biederman_fully-integrated_2013, lee_power-efficient_2015, lo_176-channel_2016, shin_flexible_2017, freeman_sub-millimeter_2017, jia_mm-sized_2018, lyu_energy-efficient_2018, khalifa_microbead_2019, burton_wireless_2020}, ultrasonic~\cite{charthad_mm-sized_2015, seo_wireless_2016, charthad_mm-sized_2018, ghanbari_sub-mm3_2019, piech_wireless_2020}, and optical~\cite{lee_250_2018, lim_019017mm2_2020} power transfer have been reported, achieving safe and reliable wireless power transfer with the size and power constraints to neural implants is still challenging. Existing technologies cannot simultaneously satisfy all the desired properties as summarized in Fig.~\ref{S1_Comp}. Radio-frequency EM waves are capable of delivering power to implants deep in the tissue~\cite{montgomery_wirelessly_2015, leung_cmos_2018}. However, they wrestle with size limitations of the receiver's antenna since efficient power delivery with electromagnetic waves requires antenna sizes comparable to the wavelength. Higher frequency RF is necessary for mm-scale implants, but suffers from higher tissue absorption~\cite{noauthor_ieee_nodate}, limiting the amount of power that can be safely delivered.
Near-field inductively coupling has been well developed in wireless power transfer~\cite{biederman_fully-integrated_2013, lee_power-efficient_2015, lo_176-channel_2016, shin_flexible_2017, freeman_sub-millimeter_2017, jia_mm-sized_2018, lyu_energy-efficient_2018, khalifa_microbead_2019, burton_wireless_2020}. It has less tissue absorption than RF because of the lower operating frequency. However, its power delivery is sensitive to perturbations in the distance and angle, especially when the coil is small.
Ultrasound is another promising method to further reduce implant size and body absorption~\cite{seo_wireless_2016, charthad_mm-sized_2018, ghanbari_sub-mm3_2019, piech_wireless_2020}. Compared to inductively coupled coils, its efficiency is more robust to the source-receiver misalignment~\cite{denisov_ultrasonic_2010}. However, it must overcome significant path loss caused by reflections at the boundaries between air, bone and tissue, which have different densities and acoustic properties. 
To alleviate the reflection between the air and the body, ultrasound gel is typically required for the transmitter.
The need for frequent replacements of the ultrasound gel and the need for the transmitter to be in contact with the skin can be inconvenient and unreliable for long-term clinical treatments. 
Further, the gel cannot mitigate the in-body path loss, which is 22-dB~cm$^{-1}$~MHz$^{-1}$ in the skull~\cite{hoskins_diagnostic_2010}. Optical power delivery is also advantageous in the miniaturization of neural implants~\cite{lee_250_2018}, \cite{lim_019017mm2_2020}, but it suffers energy loss due to scattering and may have difficulties in supporting the neural stimulation with higher power requirements. 

\begin{figure}[t]
      \centering
      \includegraphics[width=0.8\linewidth]{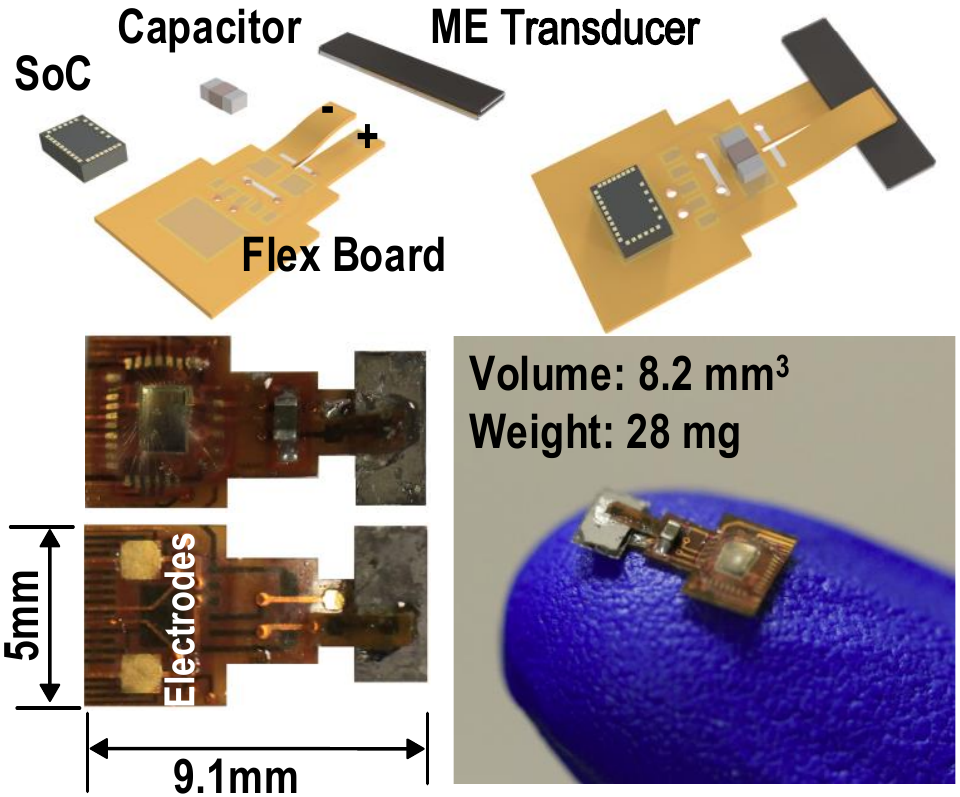}
      \caption{Illustrations of the proposed neurostimulation implant.}
      \label{S1_Device}
   \end{figure}

Among the various modalities of wireless power transfer, low-frequency magnetic field is believed to be one of the best mechanisms to safely deliver power deep inside body, because of its low absorption and strong penetration. Recently, magneothermal deep brain stimulation using magnetic nanoparticles has been demonstrated by~\cite{chen_wireless_2015}. However, the system has limited capabilities because of difficulties in the integration with other bio-electronics, especially with CMOS based devices. An alternative emerging approach to exploit low-frequency magnetic fields is by leveraging magnetoelectric (ME) effects~\cite{zaeimbashi_nanoneurorfid_2019, singer_magnetoelectric_2020}. It promises several key merits, including: (1) low tissue absorption owing to the low carrier frequency; (2) less sensitivity to changes in the alignment (in comparison with inductive coupling); and (3) high output power and efficiency with miniaturization, which will be further discussed in Section \ref{sec:ME_Power}.
The first use of ME laminates for wireless neural stimulation was recently demonstrated by Singer \textit{et al.}~\cite{singer_magnetoelectric_2020}; however, like some other analog neural stimulators \cite{chen_wireless_2015, shin_flexible_2017}, this proof-of-concept lacks robust control of the stimulation patterns and thus is highly sensitive to changes of coupling between the TX and the implants. For clinical applications, the stimulation timing and amplitude must be well controlled and programmable by the user to ensure the safety and reliability of the stimulator. Furthermore, due to the lack of energy storage and charge balancing techniques, this work has difficulties in providing large-power stimulating pulses and eliminating residual charge.
Therefore, there is a critical need to create wireless neural stimulators that simultaneously achieve clinical safety, miniaturization, operation reliability and flexibility, and programmable stimulation parameters.

To meet all the desired properties and circumvent problems mentioned above, we present MagNI (Magnetoelectric Neural Implant), the first untethered and programmable neural stimulator that exploits ME effects for power and data transfer, which integartes a 1.5~mm$^2$~180-nm CMOS system-on-chip (SoC), an in-house built 4~mm × 2~mm x 0.12~mm ME transducer, a single energy storage capacitor, and 1-mm$^2$ on-board electrodes on a flexible polyimide substrate, as shown in Fig.~\ref{S1_Device}. The proposed device features: (1) a miniature physical dimension of 8.2~mm$^3$ and 28~mg, (2) adaptive system control and data transfer mechanisms robust under source amplitude variations, (3) a 90\% chip efficiency due to its low static power down to 23.7~\si{\micro}W, and (4) the capability to perform fully programmable bi-phasic current stimulation covering 0.05 to 1.5 mA amplitude, 64 to 512~\si{\micro}s pulse width, and 0 to 200~Hz frequency ranges, making it appropriate for spinal cord stimulation to treat chronic pain.

This paper is an extended version of \cite{yu_82mm3_2020}, with more comprehensive analysis, discussions, and measurements on the safety, robustness, and power efficiency of the proposed ME power transfer mechanism. The rest of the paper is organized as follows: Section II presents qualitative and quantitative analysis of safety, misalignment sensitivity and efficiency for the ME power transfer for miniaturized neural implants; Section III describes the detailed design and implementation of the proposed SoC and neural stimulator; Section IV gives experimental results, including stimulation variability, charge imbalance, impedance of on-board electrodes, and power transfer efficiency; Section V concludes this paper.

\section{Magnetoelectric Power and Data Transfer for Neural Implants}
\label{sec:ME_Power}

\subsection{Magnetoelectric Transducers}
\label{subsec:ME_Film}

\begin{figure}[t]
      \centering
      \includegraphics[width=0.9\linewidth]{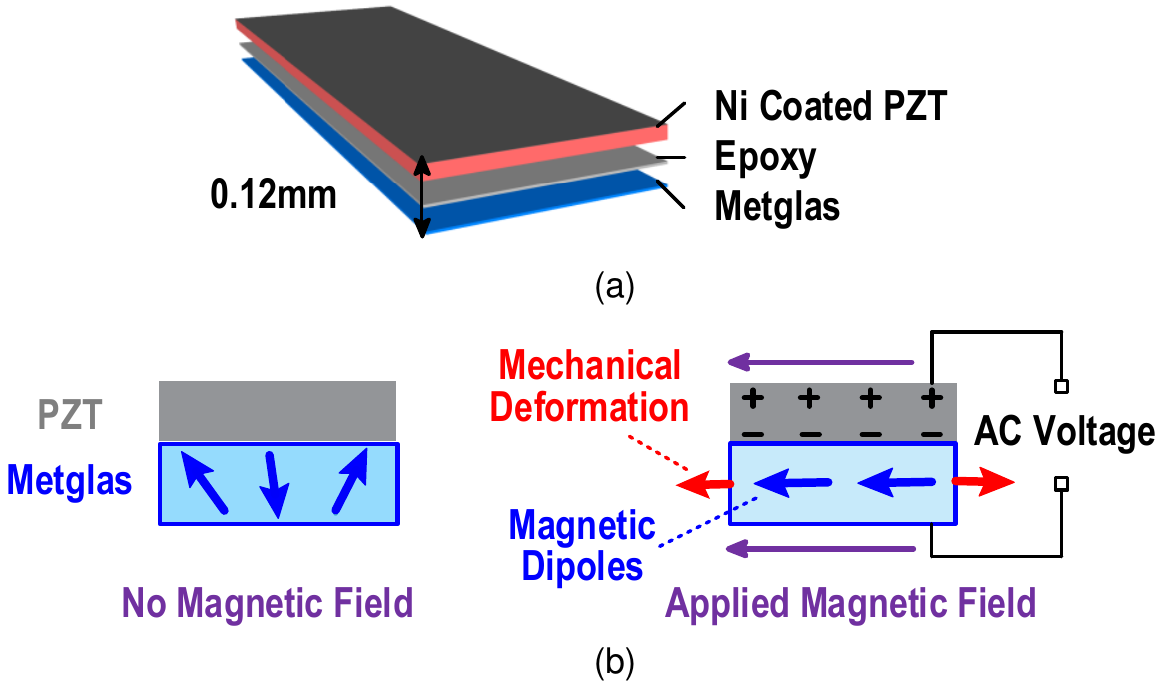}
      \caption{(a) Laminate structure and (b) operating principles of the PZT/Metglas-based magnetoelectric transducer.}
      \label{S2_Film}
   \end{figure}

\begin{figure}[t]
      \centering
      \includegraphics[width=0.9\linewidth]{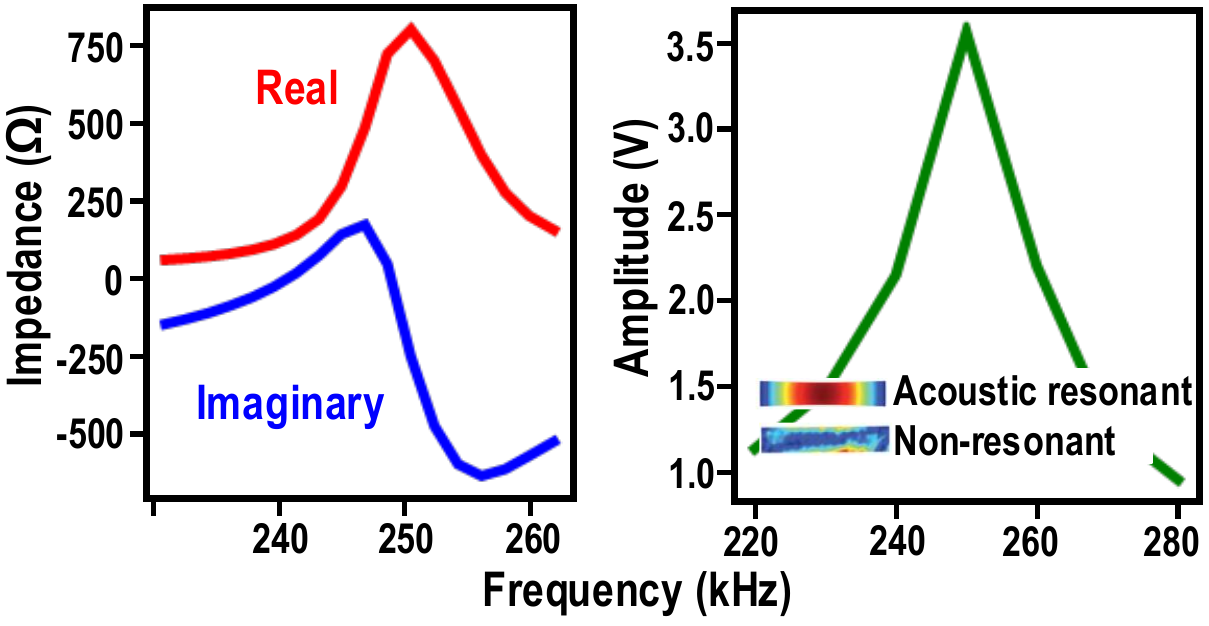}
      \caption{Measured electrical characteristics of the ME transducer.}
      \label{S2_Imp}
   \end{figure}

A magnetoelectric transducer converts an AC magnetic field to AC electrical voltage. It can be achieved by mm-scale ME laminates, which consist of a piezoelectric layer (pink), which is a nickel coated lead zirconate titanate (PZT), and a Metglas, a magnetostrictive layer (blue), as shown in Fig.~\ref{S2_Film}(a). The Metglas is made up of magnetic grains, whose domain boundaries shift with applied AC magnetic field, which causes the overall material to change  shape. The resulting vibrations are transferred to the PZT material, which develops a voltage in response to the induced strain~\cite{shuxiang_dong_longitudinal_2003, zhai_giant_2006}. Thus, a thin-film laminate converts an applied low-frequency AC magnetic field into a voltage across the transducer via mechanical coupling between magnetostrictive and piezoelectric materials (Fig.~\ref{S2_Film}(b)). The ME transducers are fabricated as laminates of Metglas (Metglas Inc.) and PZT (APC International, Ltd.). The Metglas is cut into the desired geometry from a thin sheet, while PZT is cut into the desired shape with a razor blade from a ceramic square plate. The two thin films are then coupled together with an adhesive non-conductive epoxy. Epoxy has been shown as an economical and simple solution to tightly couple the layers together for efficient energy transfer and yielding high ME voltage coefficients \cite{silva_optimization_2013}.
The resonant frequency of the ME transducers depends on the length and thickness ratio between the two laminates~\cite{dong_equivalent_2008, zhou_uniform_2014}. Thus, the specific carrier frequency of ME power link can be selected by precisely controlling the film physical dimensions.
Fig.~\ref{S2_Imp} shows that at resonance, PZT/Metglas ME films generate high output voltage (greater than 7~$\mathrm{V_\mathrm{pp}}$) with low resistive source impedance (\si{\sim}~800~\si{\Omega}), under a magnetic field of 0.57~mT, making it suitable for efficient energy harvesting.

\subsection{Safety Analysis}
\label{subsec:Safety_ME}

\begin{figure}[t]
      \centering
      \includegraphics[width=1\linewidth]{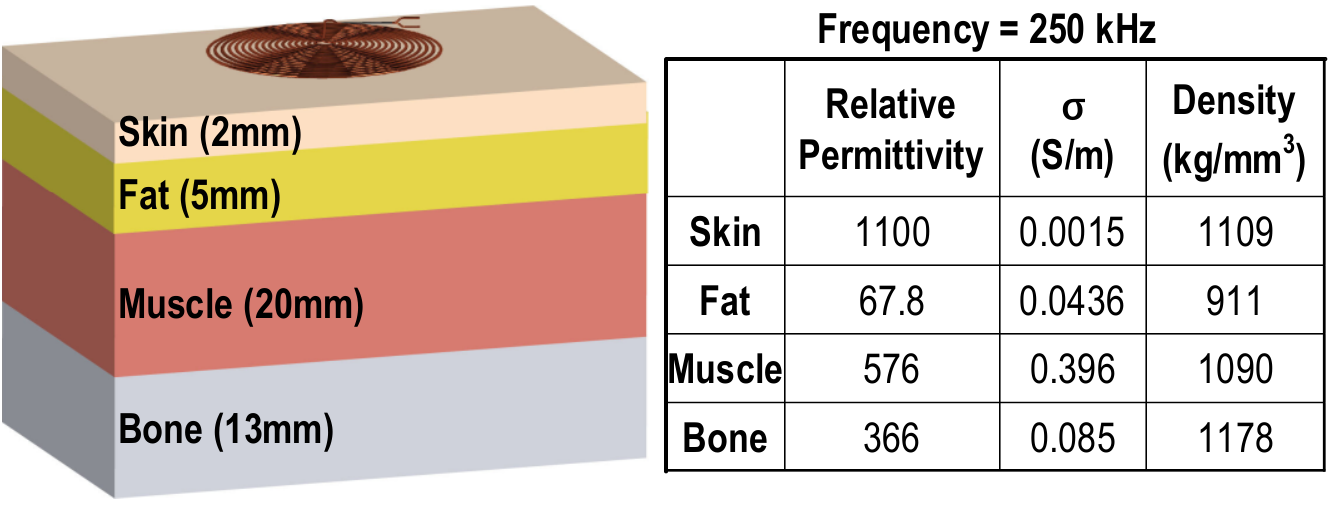}
      \caption{Multilayer human tissue model for safety analysis and material properties of each layer for the EM frequency of 250 kHz.}
      \label{S2_Model}
   \end{figure}

Body-area wireless power transfer via high frequency electric, magnetic or EM fields can cause safety issues such as heating, due to the power deposition in the human tissues. With such concerns, the field strength and frequency need to be chosen carefully to ensure safe operations in all cases.
For example, when utilizing RF field to deliver power to implants, the carrier frequencies must be at GHz to match the resonant wavelength of small antennas~\cite{montgomery_wirelessly_2015, leung_cmos_2018}, which causes high EM absorption in the body and is subject to strict safety limits. Near-field inductively powered devices mostly work at 13.56~MHz or higher frequencies to improve the quality factor of the miniaturized receiver (RX) coil for better efficiency~\cite{shin_flexible_2017, freeman_sub-millimeter_2017, jia_mm-sized_2018, lyu_energy-efficient_2018, khalifa_microbead_2019, burton_wireless_2020}, which also results in energy absorbed by the body and requires operation under certain restrictions.
The ME film works at a lower resonant frequency, which is around 250~kHz in this work.
Low-frequency magnetic fields (100~kHz to 1~MHz) can penetrate the body without substantial body absorption, resulting in alleviated safety limitations for power delivery~\cite{young_frequencydepth-penetration_1980}. 

\begin{figure}[t]
      \centering
      \includegraphics[width=1\linewidth]{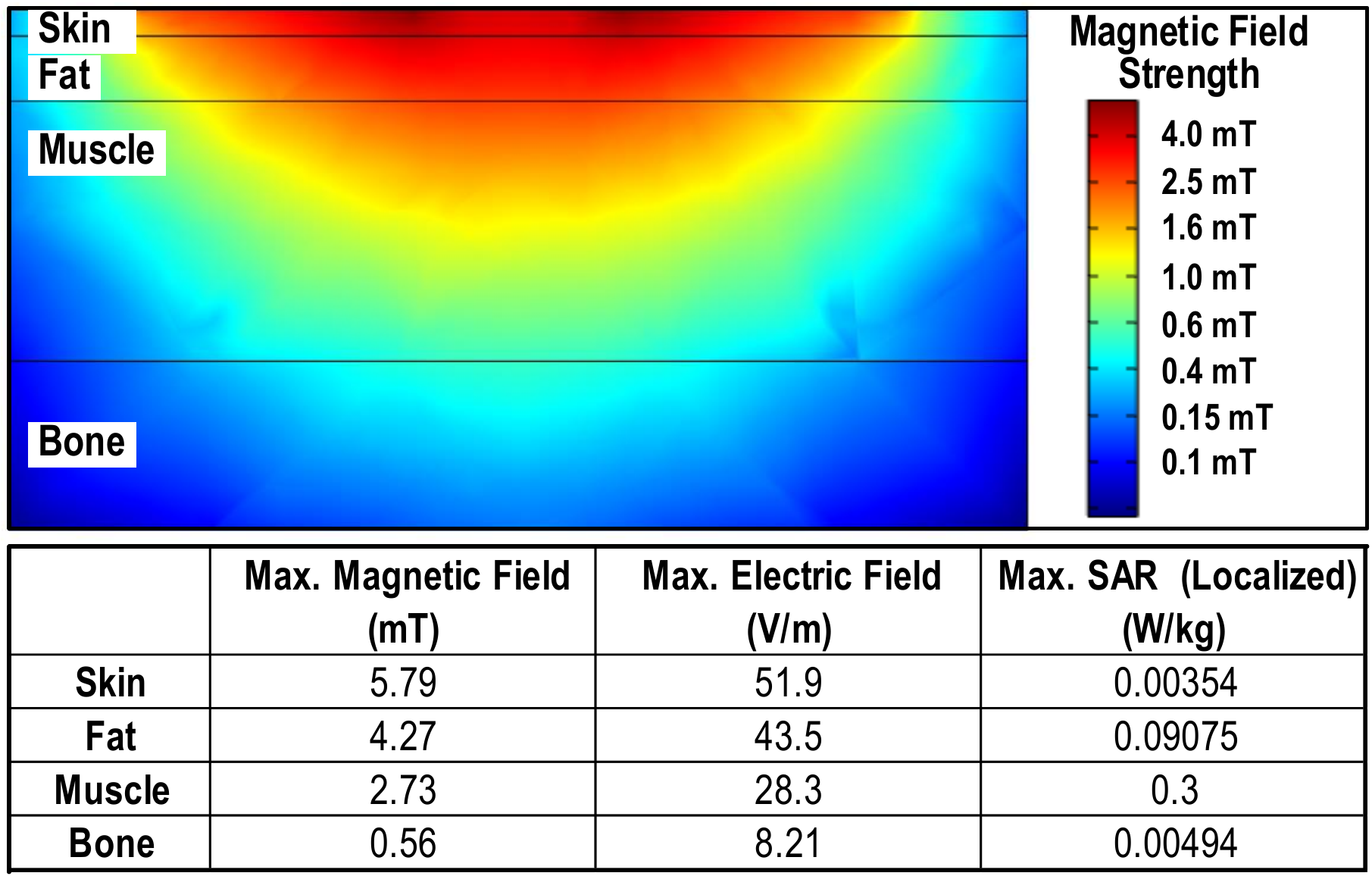}
      \caption{Local safety analysis of the 250-kHz magnetic field in COMSOL, a magnetic field strength of 0.55~mT at 30-mm depth is achieved under safety limitations.}
      \label{S2_Safe}
   \end{figure}

To assess the safety of ME power transfer for implants deep inside the body, we analyze the specific absorption rate (SAR) and induced electric fields with respect to a exposure to magnetic fields in COMSOL. The 4-layer tissue model representing spinal cord stimulation (skin, fat, muscle and bone) is built, as shown in Fig.~\ref{S2_Model}~\cite{hasgall_itis_2015}. The flat TX coil with 5.6-cm diameter, 1-mm trace width and 15 turns is employed to create the 250-kHz magnetic field. 
A 0.55-mT magnetic field strength is achievable, which is adequate for the ME film to generate a 7-$\mathrm{V_\mathrm{pp}}$ output, at a depth of 30 mm with a safe magnetic field shown in Fig. \ref{S2_Safe}. In the 4-layer tissue model, the maximum SAR is in the muscle, which is 0.3~W/kg, and the strongest electric field of 51.9~V/m is at the surface of skin, both of which satisfy the IEEE standard for humans in unrestricted environments (2~W/kg, 52.2~V/m) \cite{noauthor_ieee_nodate}. For the 250-kHz magnetic field, safety limits are dominant by the maximum allowed electric field, while the maximum SAR is still far below the limit. 

We also test the maximum allowed magnetic field strength under the SAR limitation~\cite{noauthor_ieee_nodate} for the frequency of 13.56~MHz. The comparison given in Table~\ref{table_safety} shows that the lower frequency magnetic field can be much stronger without exceeding the safety limits. Therefore, compared to inductive coupling and  RF, the ME effect has great potential to safely deliver more power to deep implants.

In addition, due to the fact that ME wireless power transfer couples magnetic fields, we wanted to assess the potential for ME power transfer to be used with magnetic resonance imaging (MRI). A study by Shellock et al., evaluated the MRI compatibility of the microstimulator RF BION, which is an inductively powered stimulator with a \si{\sim}30-mg ferrite core~\cite{shellock_implantable_2004}. The results demonstrate that patients can safely undergo MRI imaging at 1.5 T with minimal heating and artifacts being an issue only if the area of interest for imaging is proximal to the implant. In our case, Metglas will also generate artifacts; however, the amount of Metglas used in our device (\si{\sim} 0.3 mg), which is 10\si{\percent} of the amount of ferrous material in BION, will not pose as a hazard to MRI patients implanted with the device. While not fully MRI compatible as imaging artifacts may need to be considered for certain applications, ME implants should be MRI safe and clinically viable.

\begin{table}[t]
\caption{\textbf{Comparison of the Maximum Allowed Strength for 0.25 and 13.56-MHz Magnetic Fields with Safety Limitations.}}
\label{table_safety}
\centering
\setlength{\tabcolsep}{3.6pt}
\renewcommand{\arraystretch}{1.5}
\begin{tabular}{|p{55pt}|p{38pt}|p{38pt}|p{40pt}|p{38pt}|}
\hline

\centering\arraybackslash {Frequency} &
\centering\arraybackslash {Skin} &
\centering\arraybackslash {Fat} & 
\centering\arraybackslash {Muscle} & 
\centering\arraybackslash {Bone} 
\\\hline

\multicolumn{1}{|c|}{0.25 MHz}
& \multicolumn{1}{c|}{5.79 mT} 
& \multicolumn{1}{c|}{4.27 mT} 
& \multicolumn{1}{c|}{2.73 mT}
& \multicolumn{1}{c|}{0.56 mT} 
\\\hline

\multicolumn{1}{|c|}{13.56 MHz}
& \multicolumn{1}{c|}{0.35 mT} 
& \multicolumn{1}{c|}{0.26 mT} 
& \multicolumn{1}{c|}{0.17 mT}
& \multicolumn{1}{c|}{0.03 mT} 
\\\hline

\end{tabular}
\label{table_safety}
\end{table}

\subsection{Angular Misalignment Sensitivity Analysis}
\label{subsec:Angle_ME}

\begin{figure}[t]
      \centering
      \includegraphics[width=1\linewidth]{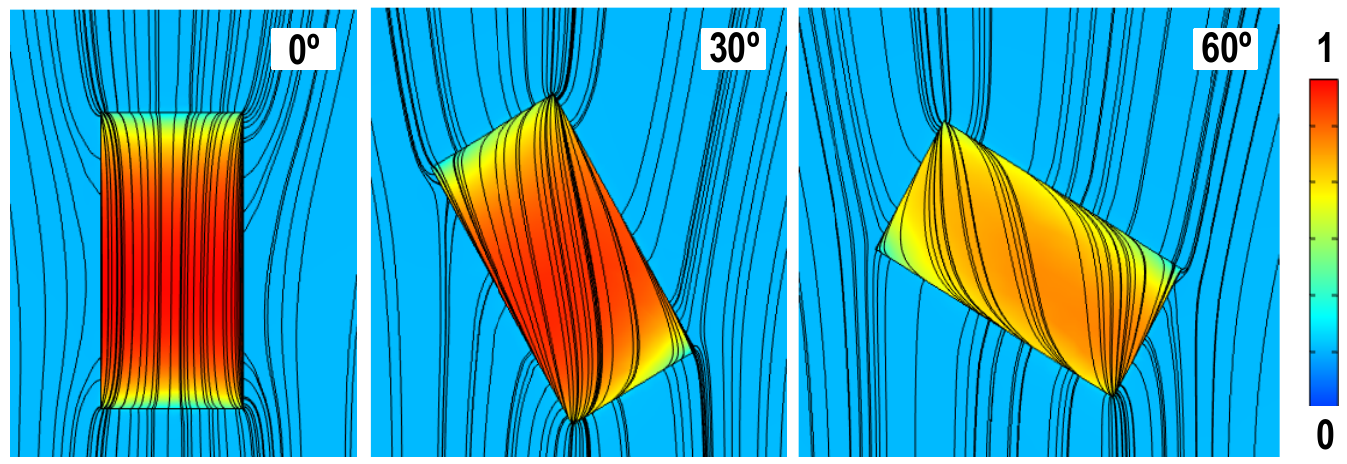}
      \caption{Simulations of magnetic flux concentration effects of the ME laminate with angle rotations, colors indicate the difference of magnetic flux density.}
      \label{S2_Flux}
   \end{figure}

Misalignment can lead to failure for most wirelessly-powered implants by negatively affecting the stability of power and data links, especially for inductively coupled approaches. The direction of the RX to the TX affects the effective area of the RX coil to catch magnetic flux, and therefore changes the output voltage and power. Theoretically, in a uniform magnetic field, the output power of the RX coil is proportional to cos$^2$(\si{\theta}), where \si{\theta} is the angle rotation.
A misaligned device is often unable to maintain correct functionality due to the insufficient levels of received power or data errors.  

ME laminates are deformed by the alternating magnetic field with the generation of stress by the magnetostrictive layer. This stress is then transferred to the PZT material, followed by crossing voltage changes, which means its output is primarily determined by the magnetic field strength instead of flux. On the contrary to inductive coils, the ideal alignment for ME is placing the laminate in parallel with magnetic induction lines. 
In addition, because of the high permeability (greater than 45000), the Metglas material has significant magnetic flux concentration effects, which enhance the magnetic flux density locally~\cite{fang_enhancing_2009}. We simulate the magnetic flux concentration effects of Metglas in COMSOL. Fig.~\ref{S2_Flux} shows that the flux density inside the laminate is much higher than that in the free space even with angle rotations, which means the decrease in flux density due to misalignment can be partially compensated by the magnetic flux concentration effect. As a result of these unique characteristics, compared to the inductive coil, the ME transducer is less sensitive to angular misalignment. 

\begin{figure}[t]
      \centering
      \includegraphics[width=0.9\linewidth]{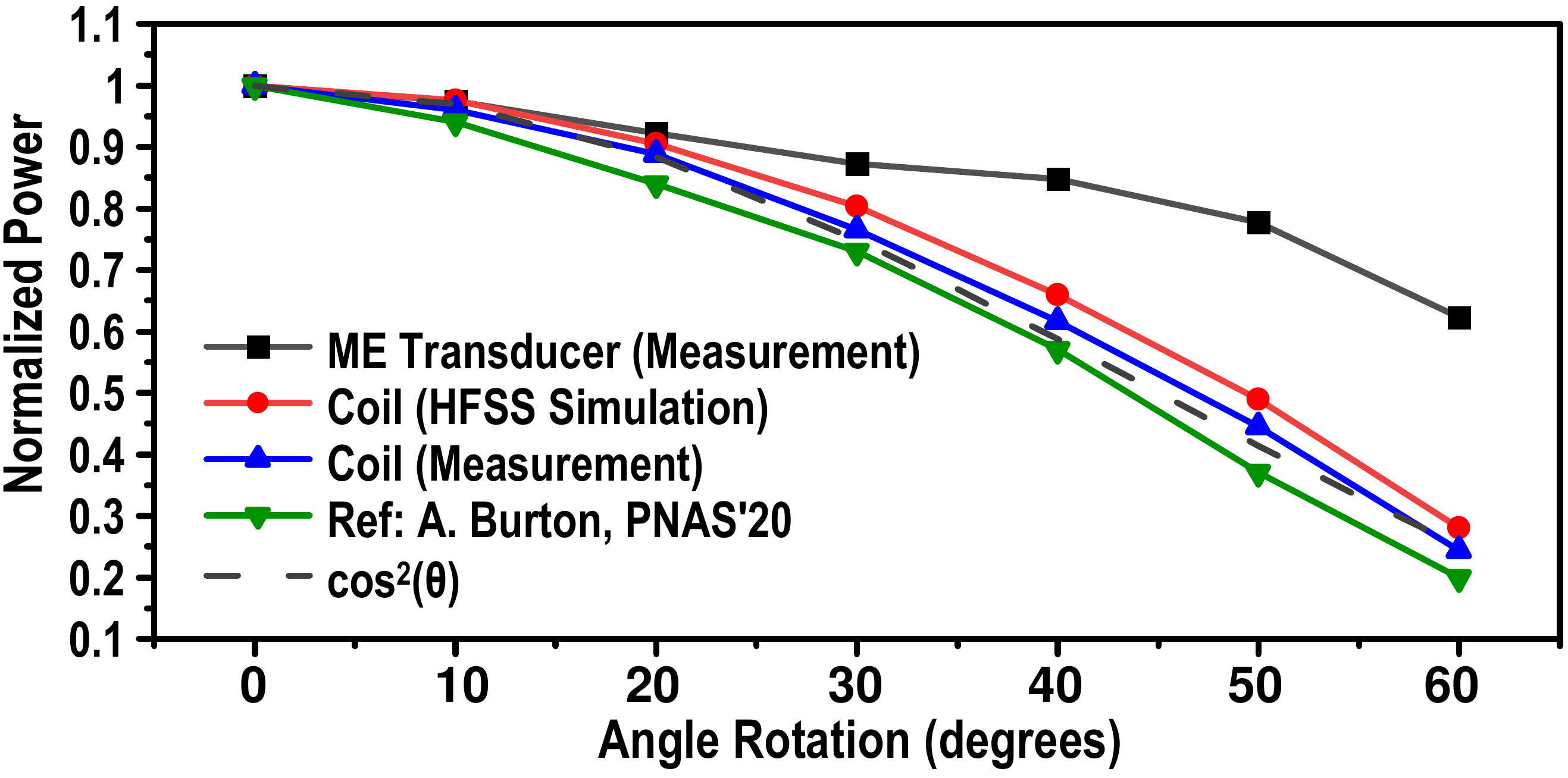}
      \caption{Sensitivity tests of angular misalignment of ME transducers, in comparison with inductive coils \cite{burton_wireless_2020}}
      \label{S2_Angle}
   \end{figure}
   
   \begin{figure}[t]
      \centering
      \includegraphics[width=1\linewidth]{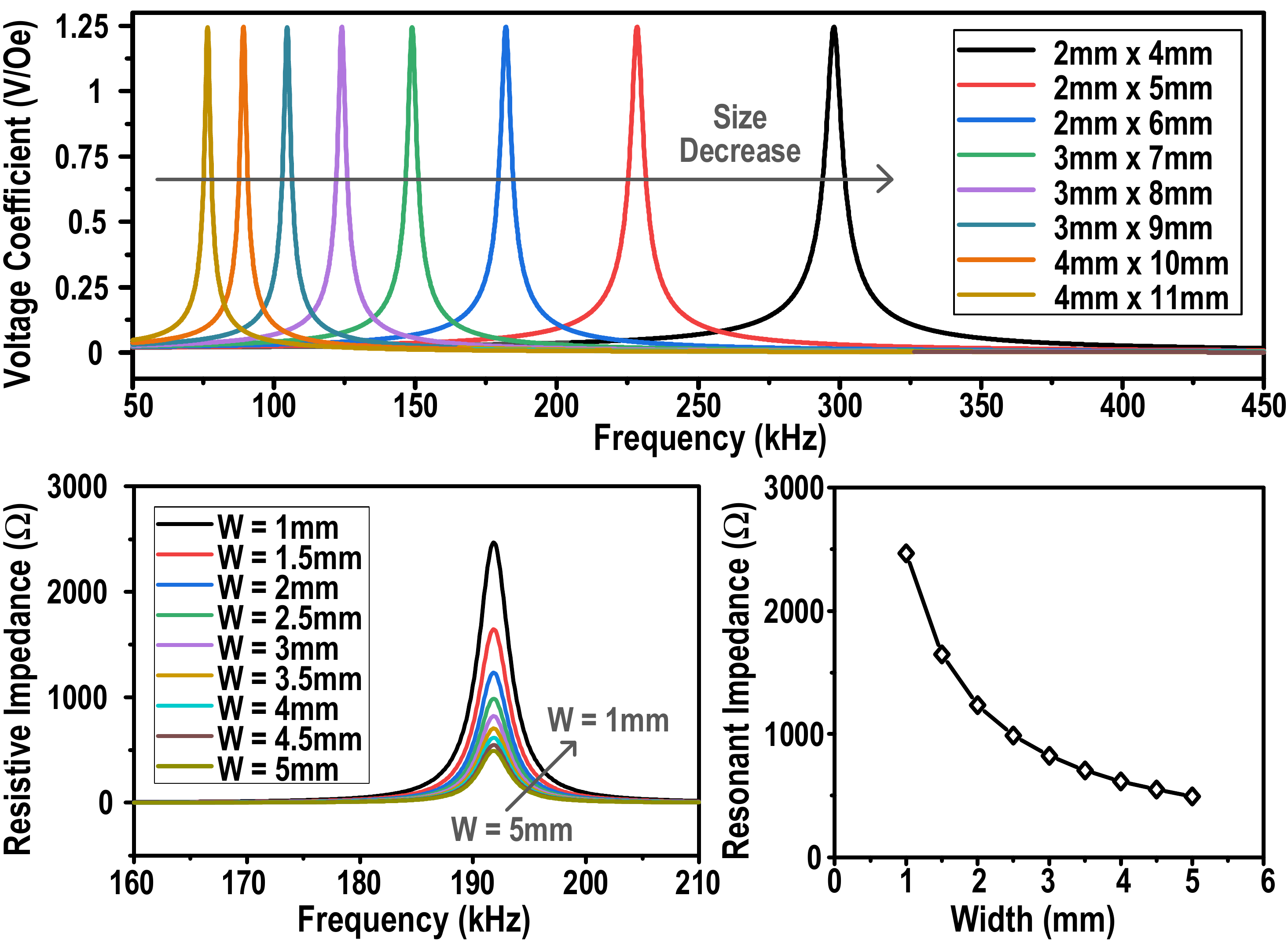}
      \caption{Simulations of the voltage coefficient of ME transducers with different sizes (W x L) and simulations of the impedance at resonance of ME transducers with different widths (L = 6~mm).}
      \label{S2_Sim_ME}
   \end{figure}

To test the angle dependence of the ME, we measure 12-mm$^2$ ME transducers with a 25-mm cylindrical TX coil with angle rotations. For comparison, both simulations and measurements of inductive coils are conducted with the same TX but operating at 13.56~MHz. RX coils used for the experimental test are built in boards with a diameter of 4~mm, the number of turns of 10 and a quality factor of 30.6. The simulated and experimental results of the inductive coil match well with the theoretical value. Measurement results reported by~\cite{burton_wireless_2020} are also added as a reference, whose RX coil is 10.5 x 7~mm$^2$, 7 turns with a Q of 23.05. Fig.~\ref{S2_Angle} shows a 3X maximum improvement of ME in angular misalignment robustness compared to inductive coupling.

\begin{figure}[t]
      \centering
      \includegraphics[width=0.9\linewidth]{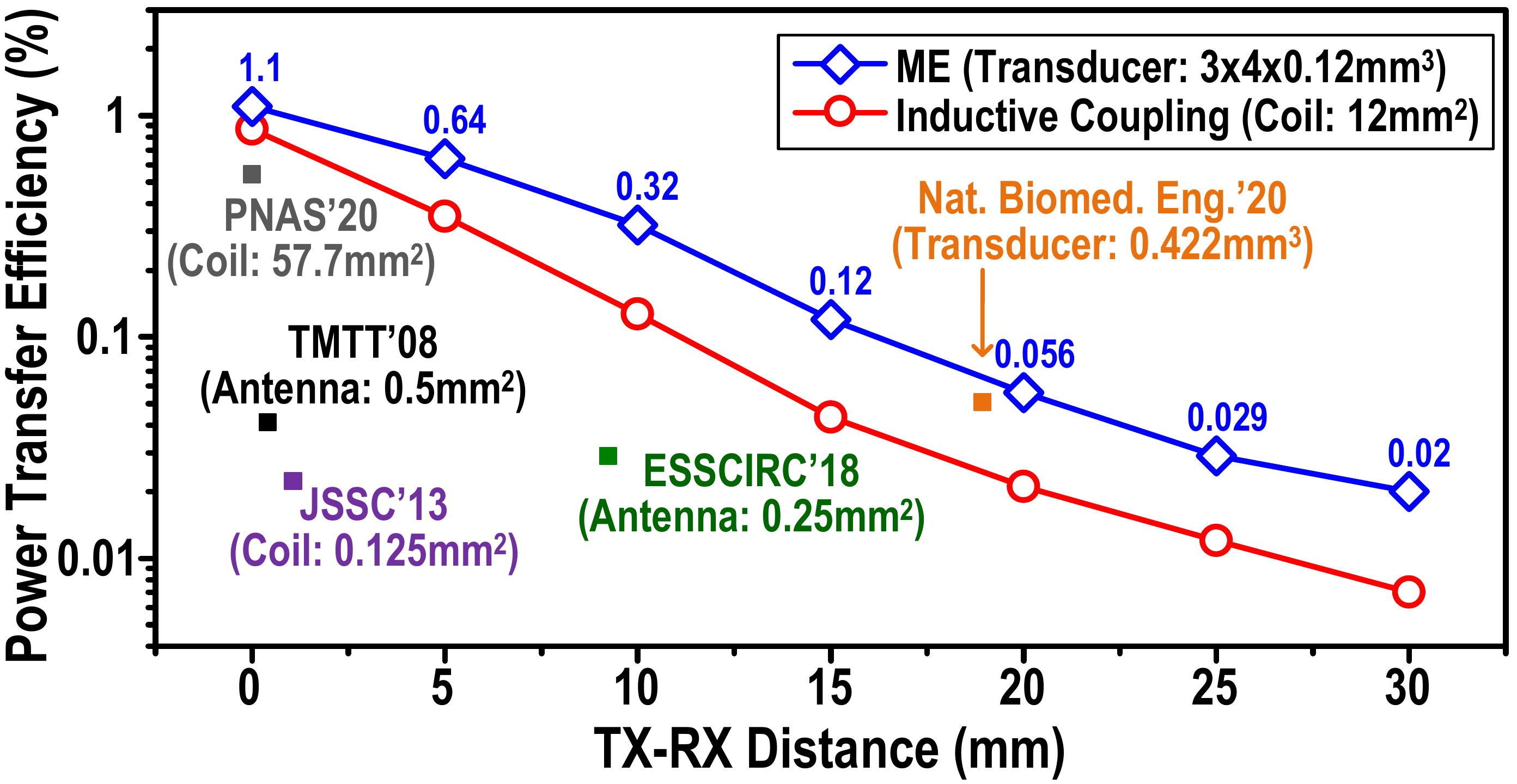}
      \caption{Measurements of ME power transfer efficiencies over various TX-RX distances, in comparison with RF, inductive coupling and ultrasonic power transfer~\cite{biederman_fully-integrated_2013, burton_wireless_2020, chen_245-ghz_2008, leung_cmos_2018, piech_wireless_2020}.}
      \label{S2_Eff}
   \end{figure}
   
   \begin{figure}[t]
      \centering
      \includegraphics[width=1\linewidth]{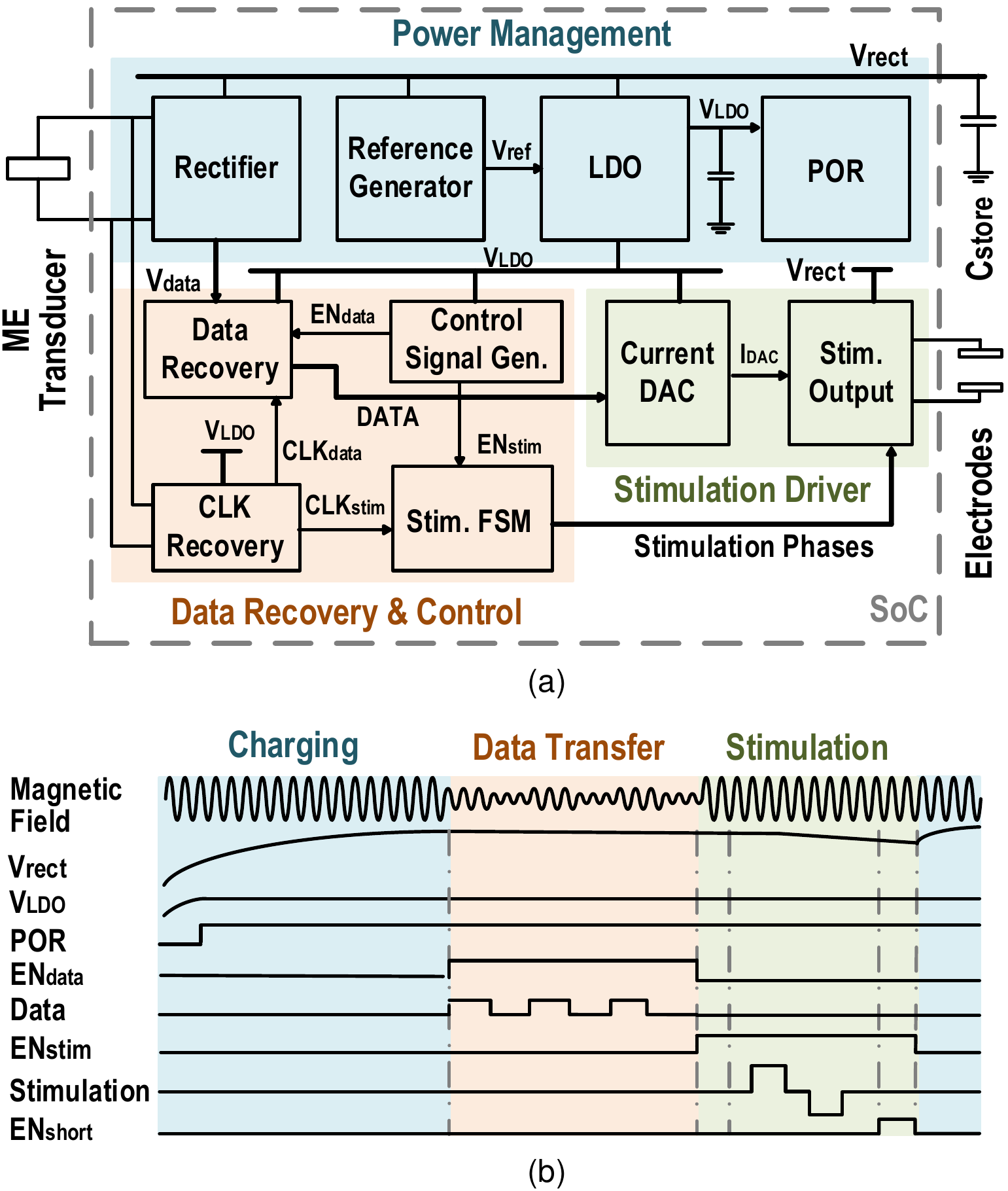}
      \caption{(a) System diagram and (b) operation waveforms of the MagNI SoC.}
      \label{S3_SoC}
   \end{figure}

\subsection{Power and Efficiency Analysis}
\label{subsec:Eff_ME}
   
Miniaturization of the wireless powered implant can influence the overall power efficiency of the link. With far-field power transfer taking advantage of EM wave radiations, the received power is quite weak when the antenna is smaller than one-tenth of the wavelength \cite{landgren_wideband_2017}. As another example, in the near field, the voltage and power of the RX coil is also highly dependent on the size. Faraday's Law gives the induced voltage of the coil as
\begin{equation}
\begin{split}
V_\mathrm{coil,RX} 
 =  N \frac{d\si{\Phi}}{dt} = A\times N\frac{dB}{dt},
\end{split}
\label{Vcoil}
\end{equation}
where A is the effective area of the coil, B is the magnetic flux density and N is the number of turns. By assuming that the coil is flat and circular, an estimation of the output power can be given as    
\begin{equation}
\begin{split}
P_\mathrm{coil,RX}\ 
 \si{\approx}\ \frac{\si{\pi}^2 N B R^3}{4 \si{\micro}} \times \frac{dB}{dt},
\end{split}
\label{Pcoil}
\end{equation}
where R is the diameter of the coil and \si{\micro} is the permeability. Shrinking down the coil size not only reduces the amount of captured flux, but also degrades the quality factor, which causes a significant decrease in the received power.

Previous studies~\cite{dong_equivalent_2008, zhou_uniform_2014, truong_experimentally_2020} developing the equivalent circuit model of ME laminated composites have shown that its voltage coefficient \si{\alpha} (defined by dV/dH with units of V/Oe) is independent of width and length, which means scaling down the ME laminate area does not reduce the output voltage. However, the resonant frequency is inversely proportional to the length while the source impedance at resonance is inversely proportional to the width. As a result, the theoretical maximum output power of the ME laminate is linearly related to its width. Interestingly, in comparison to inductive coupling, the ME received power demonstrates less dependency on the transducer size, which demonstrates the potential for superior scalability in miniaturizing the device. These phenomena are verified by simulation results given in Fig.~\ref{S2_Sim_ME} and measurements of~\cite{singer_magnetoelectric_2020}.

To evaluate the efficiency of ME against inductive coupling, both of which are based on low-frequency magnetic fields, we measure the ME and inductive coupling power transfer efficiency (PTE) across TX-RX distances in air (Fig.~\ref{S2_Eff}) using custom built 12-mm$^2$ ME transducers and 12-mm$^2$ 10-turn coils on PCB. As expected, PTEs of both modalities decrease as the distance increases, but ME consistently presents a higher PTE across the range and a maximum efficiency of 1.1\% is achieved when the ME transducer is at the center of the TX coil. Additionally, reported PTE of various miniaturized devices ~\cite{biederman_fully-integrated_2013, burton_wireless_2020, kuo_equation-based_2018, leung_cmos_2018, piech_wireless_2020} are included in Fig.~\ref{S2_Eff} for comparison. Among them, \cite{ chen_245-ghz_2008, leung_cmos_2018} utilize RF, \cite{ burton_wireless_2020, biederman_fully-integrated_2013} adopt inductive coupling power transfer and~\cite{piech_wireless_2020} establishes the power link by ultrasound. 
It is worth noting that the decrease of PTE at larger TX-RX distances is solely caused by the decline of magnetic field strength, the ME energy conversion efficiency (magnetic to mechanical, mechanical to electrical) that relies on the coupling between laminates depends solely on intrinsic material properties and thereby remains the same at varying distances \cite{truong_fundamental_2020}.

\begin{figure}[t]
      \centering
      \includegraphics[width=0.95\linewidth]{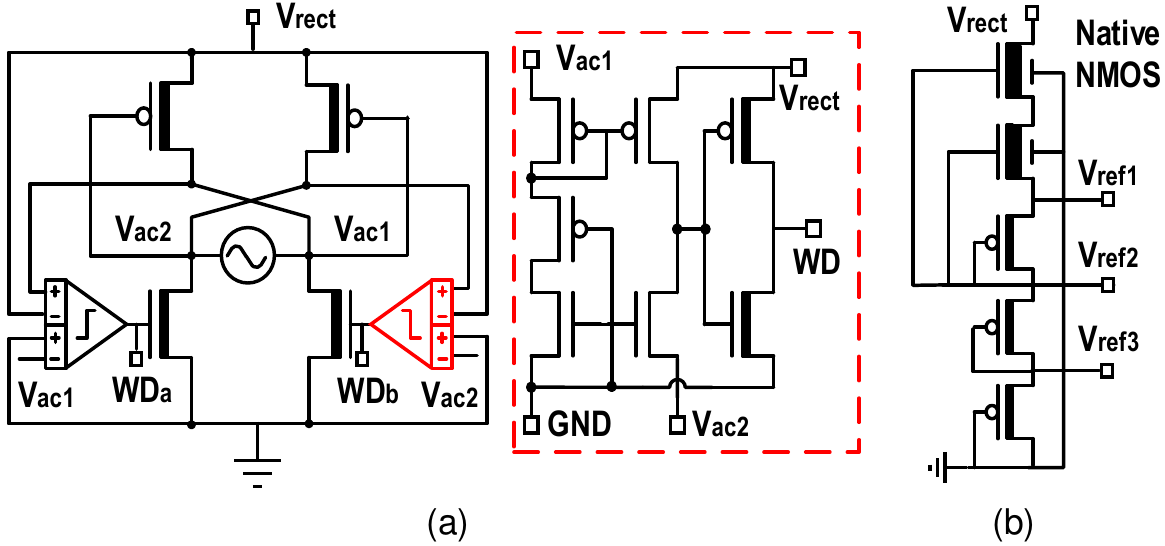}
     \caption{Schematics of (a) the active rectifier with 4-input comparator and (b) the native MOS-based voltage reference generator.}
      \label{S3_Rec}
 \end{figure}

\section{MagNI System-on-Chip Design}
\label{sec:magni_soc_design}

\subsection{System Overview}
\label{subsec:SoC_Sys}

 
The MagNI SoC, consisting of power management, data recovery, and stimulation modules, interfaces with a ME film to receive power and data and drives programmable stimulation, as shown in Fig. \ref{S3_SoC}(a). The SoC cycles through charging, data receiving, and stimulation phases at the desired stimulation frequency set by the external transmitter (Fig. \ref{S3_SoC}(b)).

\subsubsection{Power Management}
AC voltage induced on the ME transducer is first converted to a DC voltage ($V_\mathrm{rect}$) by the active rectifier (Fig. \ref{S3_Rec}(a))~\cite{lam_integrated_2006}. Energy is then stored in an off-chip 4.7-\si{\micro}F capacitor for the high-power stimulation. The low-dropout regulator (LDO) provides a constant 1-V supply $V_\mathrm{LDO}$ for digital control and data transfer to reduce power consumption. Temperature and supply-invariant reference voltages for the entire system are generated on chip by a reference circuit with native NMOS and stacked diode-connected PMOS transistors~\cite{seok_portable_2012}. Three different voltage references of 1 V, 0.6 V and 0.3 V are provided by the ultra low power circuit shown in Fig. \ref{S3_Rec}(b). 

 \begin{figure}[t]
      \centering
      \includegraphics[width=1\linewidth]{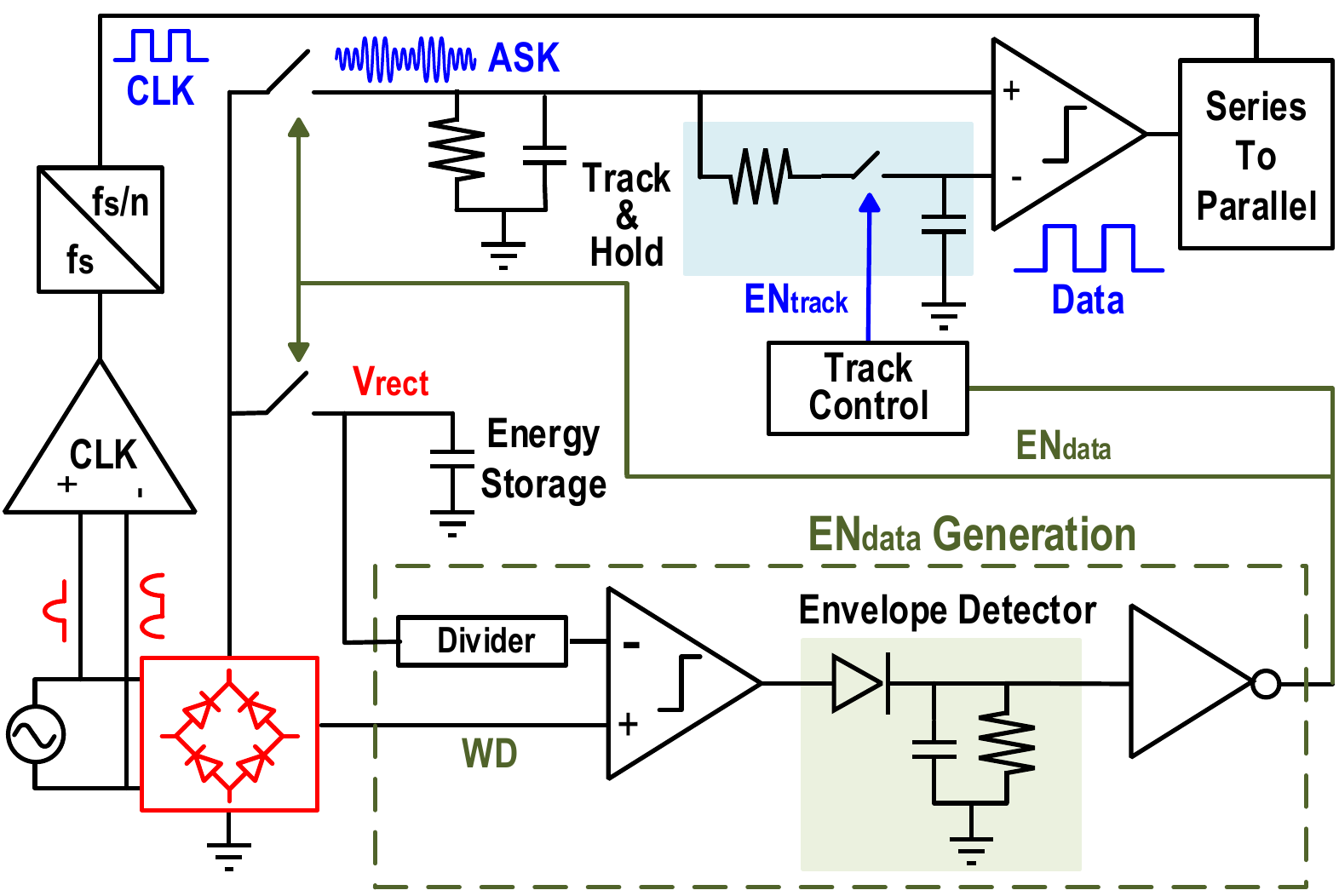}
      \caption{Schematic of the proposed adaptive operating scheme.}
      \label{S3_Control}
   \end{figure}
  
 \begin{figure}[t]
      \centering
      \includegraphics[width=0.85\linewidth]{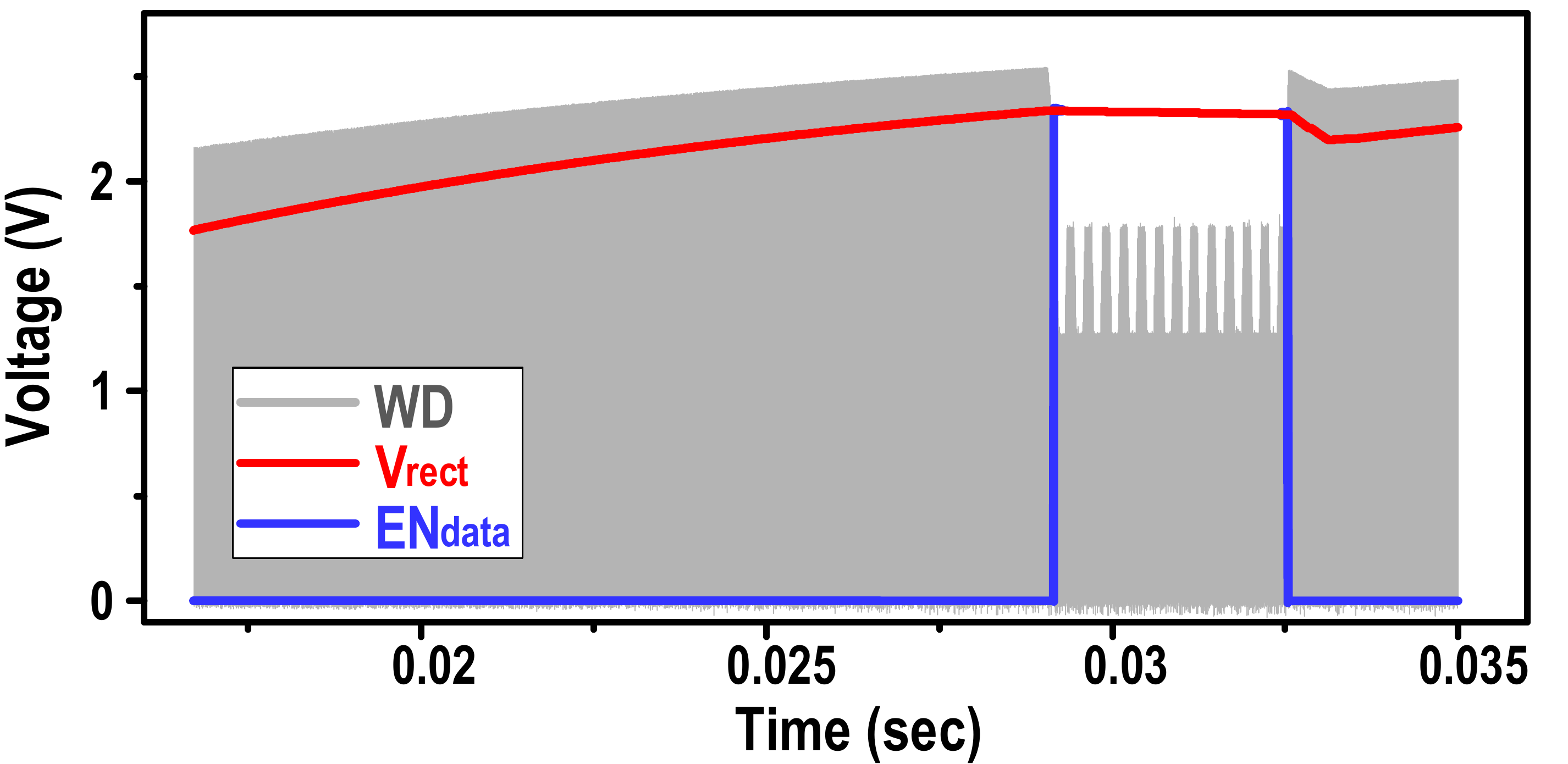}
      \caption{Simulated waveform of $EN_\mathrm{data}$ generation for operation phases control.}
      \label{S3_ENdata}
   \end{figure}
   
   \begin{figure}[t]
      \centering
      \includegraphics[width=0.9\linewidth]{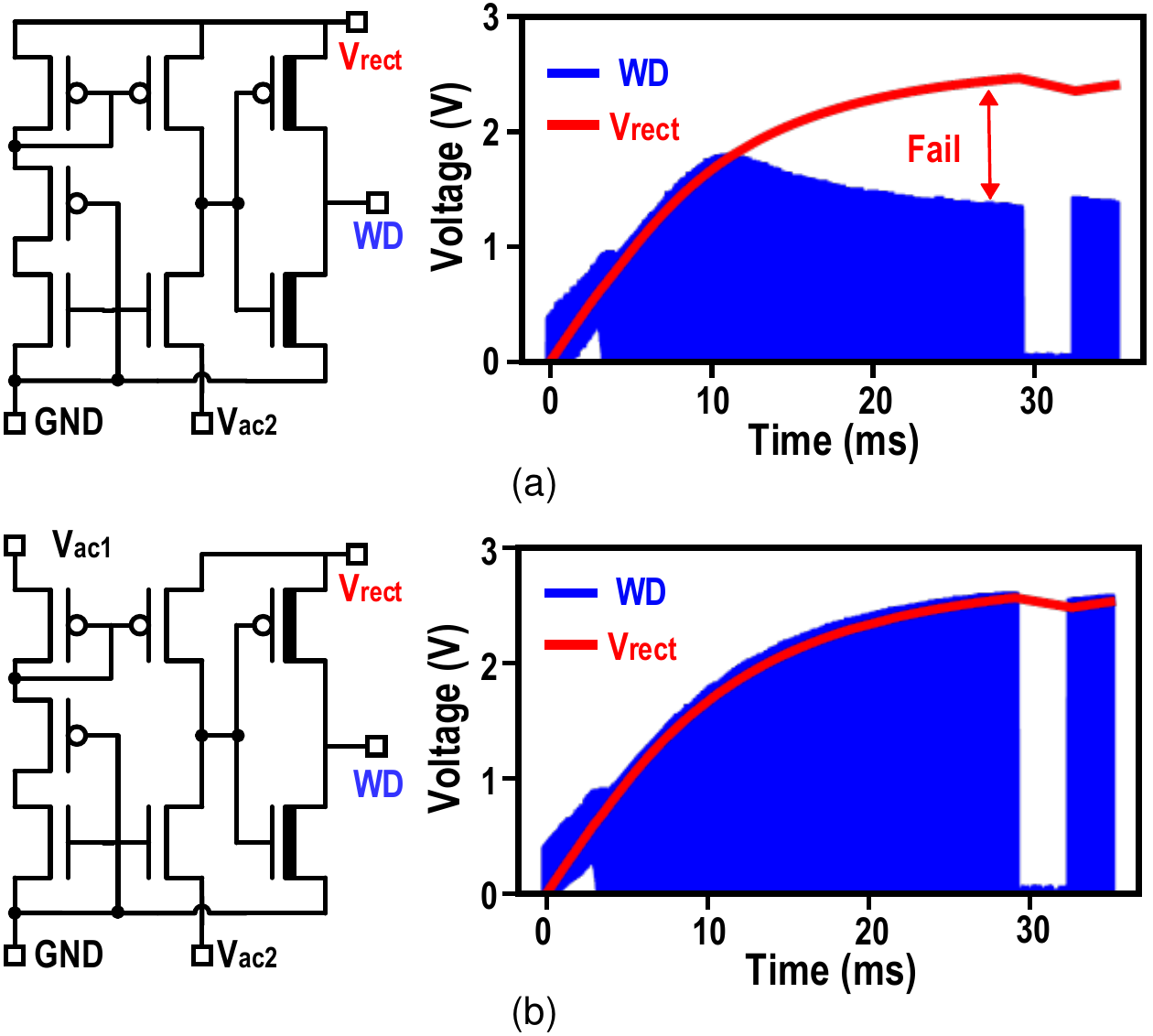}
      \caption{(a) Schematic of the 2-input compactor of active rectifier and its simulated output WD with 23-mV offset; (b) Schematic of the 4-input compactor of active rectifier and its simulated output WD with 23-mV offset; the WD generated by 2-input comparator fails in tracking $V_\mathrm{rect}$.}
      \label{S3_WD}
   \end{figure}

\subsubsection{Data Recovery}
Downlink data is transmitted by amplitude shift keying (ASK) modulation of the magnetic field and is used to program stimulation patterns. To save energy, we reuse the rectifier to detect data amplitudes. However, the large energy storage capacitor prohibits the rectified voltage change in a short time. To solve this problem, a dual path architecture is adopted, which has an auxiliary path enabled by a global control signal $EN_\mathrm{data}$ for data communication.

\subsubsection{Stimulation}
The 5-bit current DAC, which is only enabled in the stimulation phase to save power, programs stimulation amplitudes with a 50-\si{\micro}A resolution. A finite-state machine (FSM), also configured by the received data, controls the stimulation timing. Its operating phases include initialization, anodic stimulus, inter-phasic pause, cathodic stimulus and electrodes shorting. Bi-phasic stimulation, unlike a mono-phasic pulse, is charge-balanced and prevents undesired electrochemical reactions on electrodes. It is realized by the H-bridge based stimulation driver controlled by $\si{\Phi}_\mathrm{1}$ and $\si{\Phi}_\mathrm{2}$. The inter-phase pause with a fixed duration of 32 \si{\micro}s allows the stimulation driver to stabilize after the phase switching and reduces the threshold for bi-phasic stimulation \cite{gorman_effect_1983}.  By the end of stimulation, electrodes are shorted by $EN_\mathrm{short}$ to completely remove the residual charge.

\subsection{Design Considerations for Robust Operation}
\label{subsec:SoC_Control}

In order to maintain system operation robustness under varying conditions, such as amplitude changes due to varying distances and misalignment between device and transmitter, we make the device operation adaptive, calibration free, and fully controllable by the TX. Fig.~\ref{S3_Control} shows the proposed architecture to realize the robust operating scheme.

First, the enabling of the data transfer phase is controlled by changing the ME amplitude, so that scheduling of the implant is fully controlled by the external TX with accurate timing references and computation resources. Because the large energy storage capacitor prevents $V_\mathrm{rect}$ to quickly change, the existing comparator in the active rectifier is employed as a watchdog to monitor the ME induced voltage change. It generates a train of pulses WD tracking $V_\mathrm{rect}$ during
charging phase and stops once the input’s amplitude is below the voltage of the energy capacitor. By comparing the watchdog signal with divided $V_\mathrm{rect}$ and
extracting the envelop, a rail-to-rail start signal for data transfer ($EN_\mathrm{data}$) is created, as shown in Fig.~\ref{S3_ENdata}. Comparator offset may cause a failure in tracking $V_\mathrm{rect}$ and generate incorrect $EN_\mathrm{data}$ (Fig.~\ref{S3_WD}(a)). To alleviate this, 4-input comparator~\cite{lam_integrated_2006} is employed to double the sensing margin and increase offset tolerance from 11mV to 23mV (Fig.~\ref{S3_WD}(b)), effectively reducing the failure probability from 8\% to 0.02\% (calculated with simulated variance of offset). 

\begin{figure*}[t]
      \centering
      \includegraphics[width=0.9\linewidth]{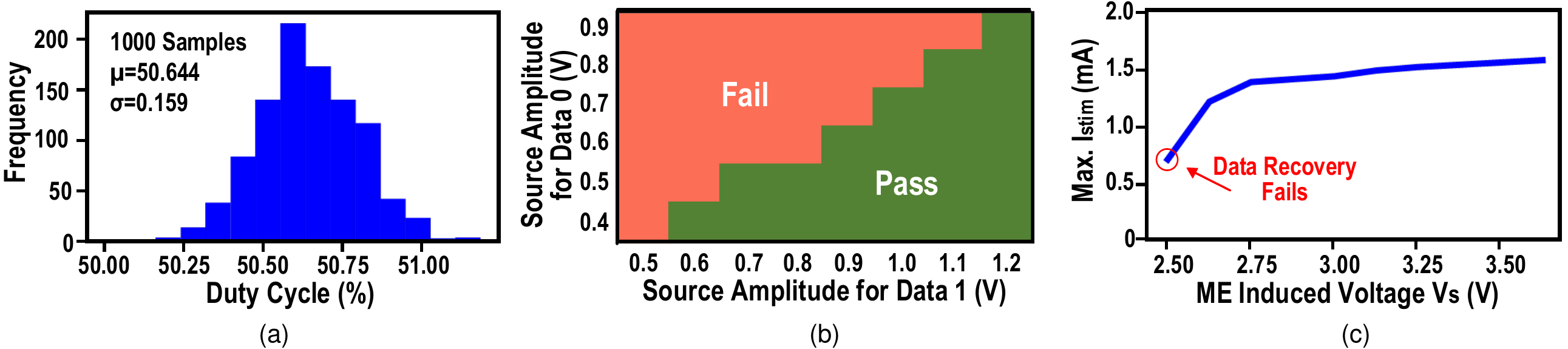}
      \caption{(a) Monte Carlo simulations of the duty cycle of the recovered clock; (b) Measured shmoo plot for data recovery with different source amplitudes; and (c) stimulation current measurement under varying ME induce voltage.}
      \label{S3_Data}
   \end{figure*}

Second, the system global clock is derived directly from the ME source by a low-power comparator, and therefore presents a process, supply and temperature-invariant frequency. The Monte Carlo simulation result given in Fig.~\ref{S3_Data}(a) demonstrates a tight distribution of duty cycle with \si{\sigma} of 0.159\%. 

   \begin{figure}[t]
      \centering
      \includegraphics[scale=0.35]{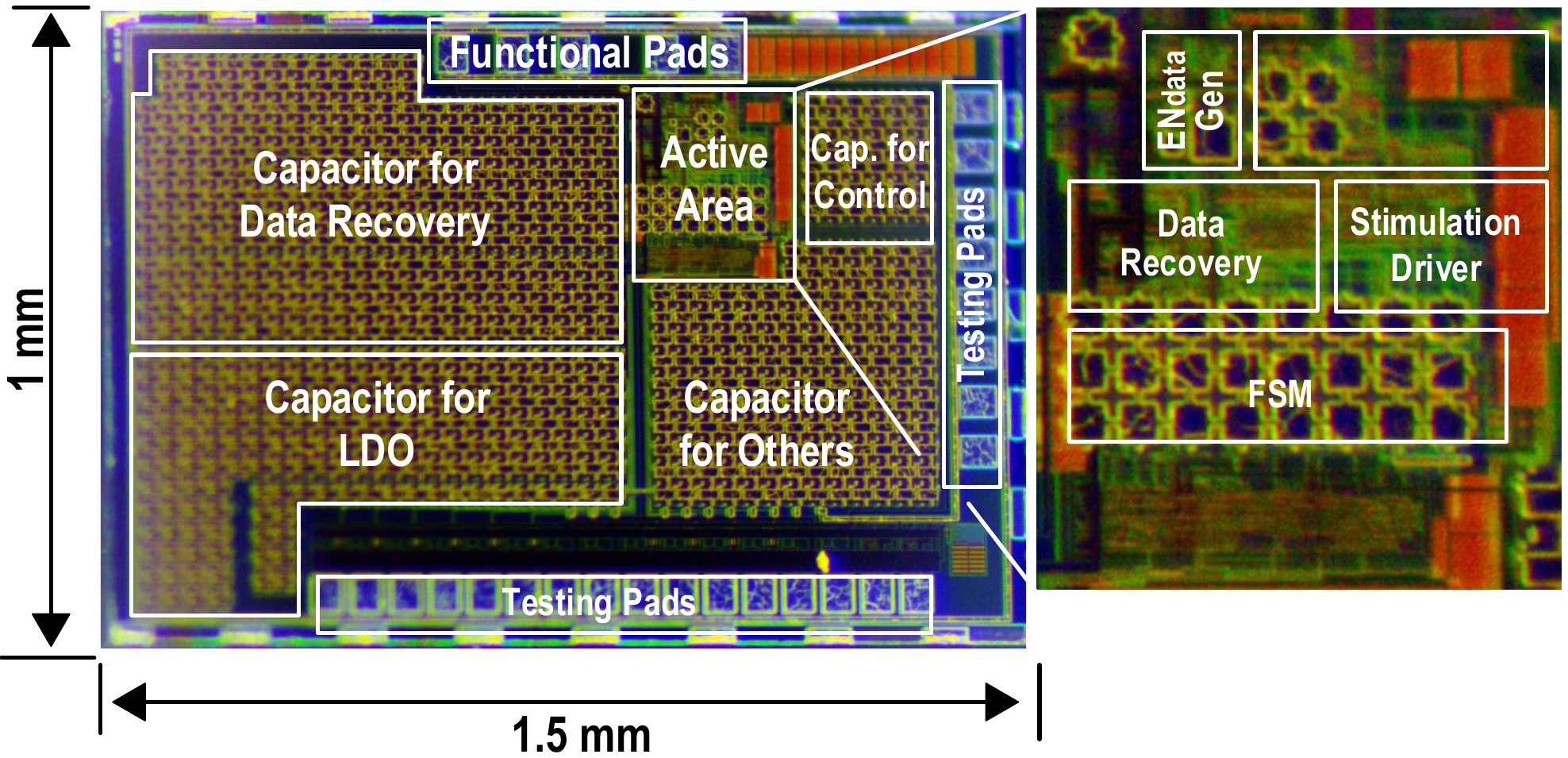}
      \caption{Micrograph of the MagNI SoC.}
      \label{S4_Die}
   \end{figure}
   
\begin{figure}[t]
      \centering
      \includegraphics[scale=0.55]{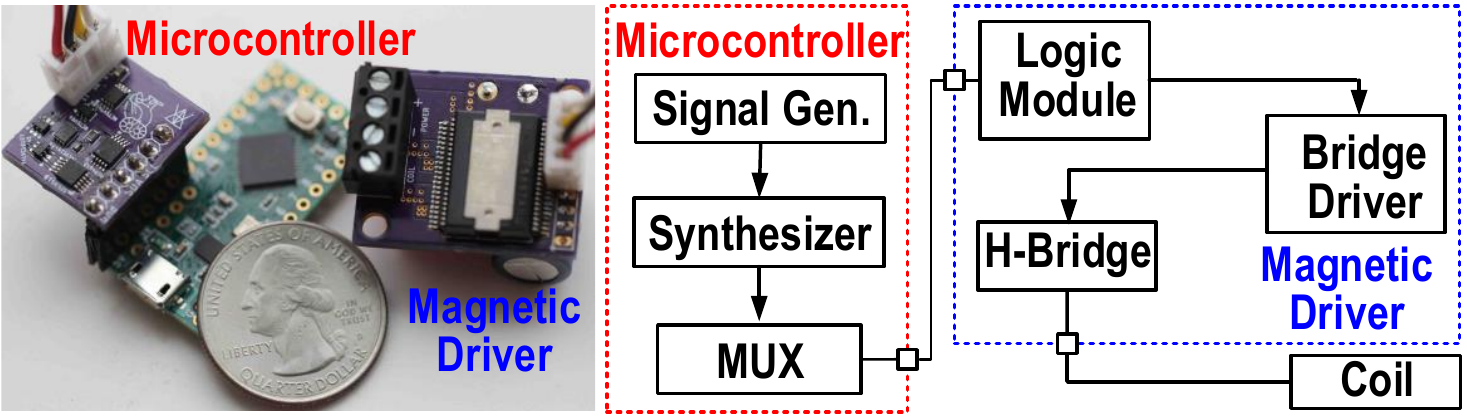}
      \caption{Illustration of the portable TX consisting of microcontroller and magnetic driver.}
      \label{S4_TX}
   \end{figure}

Third, to ensure correct data demodulation under ME voltage variations, the voltage threshold is generated online before every data transfer, using an alternating pilot tone sent by
the TX. The threshold is extracted with a low-pass filter followed by track and hold circuits. Measurement results show successful data recovery without errors with a wide source amplitude range (minimum of 0.6 V for data “1” and minimum of 0.4 V for data “0”) and 0.2-V minimum amplitude difference (Fig.~\ref{S3_Data}(b)). 
Furthermore, we validated the robust operation by showing that a fixed setting of the stimulation current stays around the expected 1.5~mA, when the ME-induced voltage varies from 2.6 to 3.6~V due to magnetic field variations (Fig.~\ref{S3_Data}(c)). Maximum $I_\mathrm{stim}$ falls slightly with the decrease of source amplitude reducing voltage headroom, until data recovery fails when the $V_\mathrm{s}$ is below 2.5~V.

\section{Experimental Results}
\label{sec:measurement_result}

\subsection{Electrical Validation of the MagNI SoC}
\label{subsec:F_V_SoC}

\begin{figure}[t]
      \centering
      \includegraphics[scale=0.58]{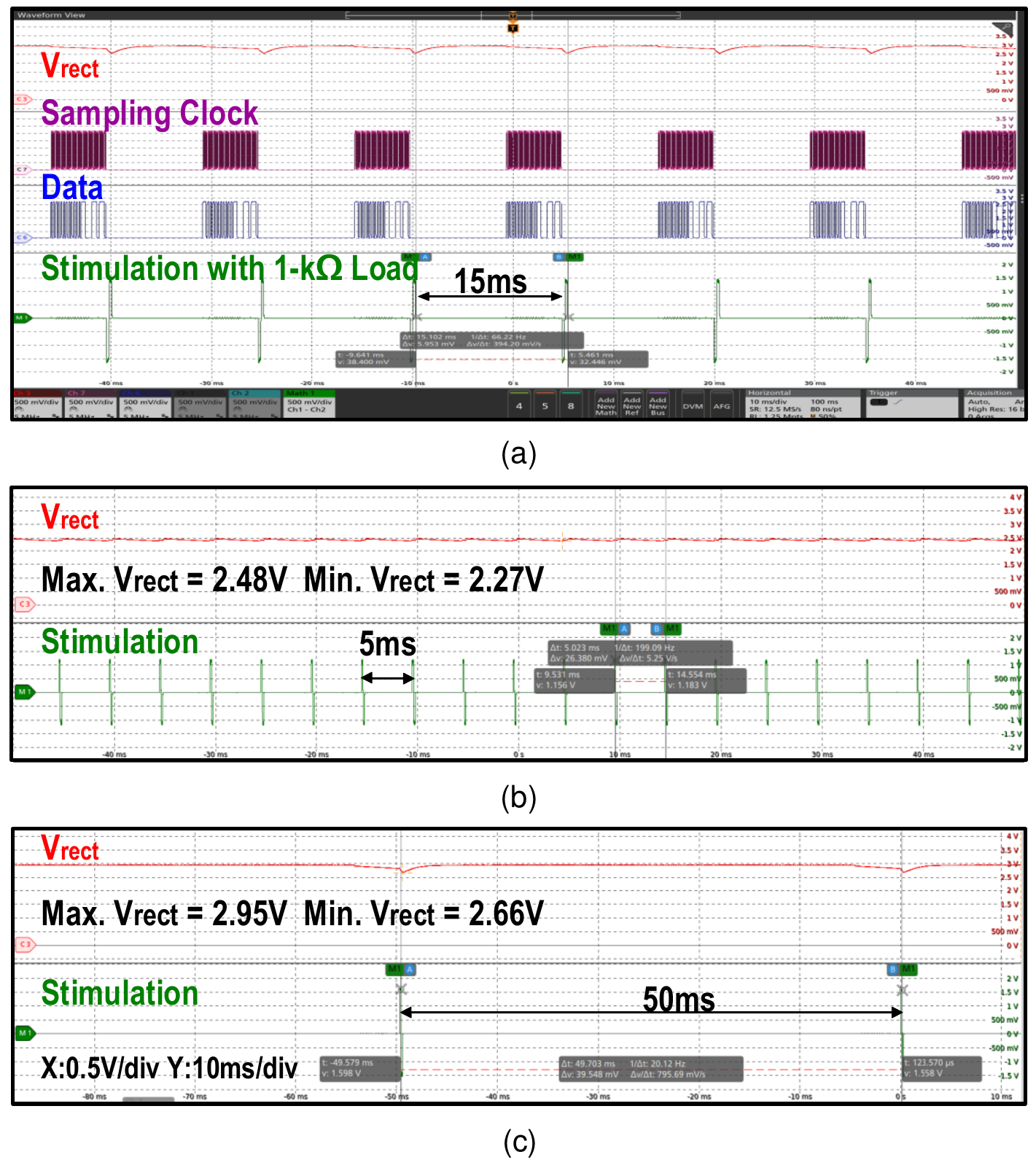}
      \caption{Measured waveforms of MagNI operating at stimulation frequencies of (a) 67~Hz, (b) 20~Hz and (c) 200~Hz.}
      \label{S4_Waveform}
   \end{figure}
   
\begin{figure}[t!]
      \centering
      \includegraphics[scale=0.7]{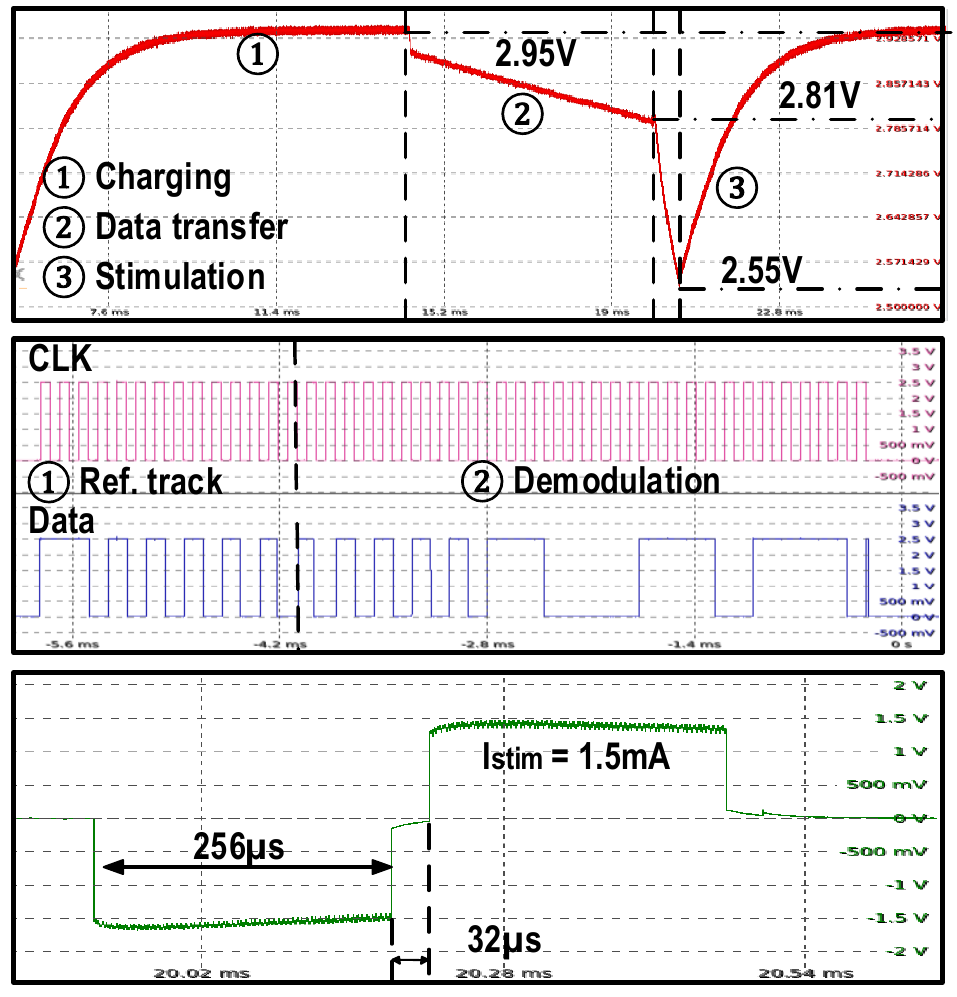}
      \caption{Measurements of rectified voltage at different operating phases; clock and data recovery for ASK modulation; and 1.5-mA, 512-\si{\micro}s bi-phasic current stimulation, at 67-Hz stimulation frequency.}
      \label{S4_Zoomin_2}
   \end{figure}
   
   \begin{figure}[t!]
      \centering
      \includegraphics[scale=0.28]{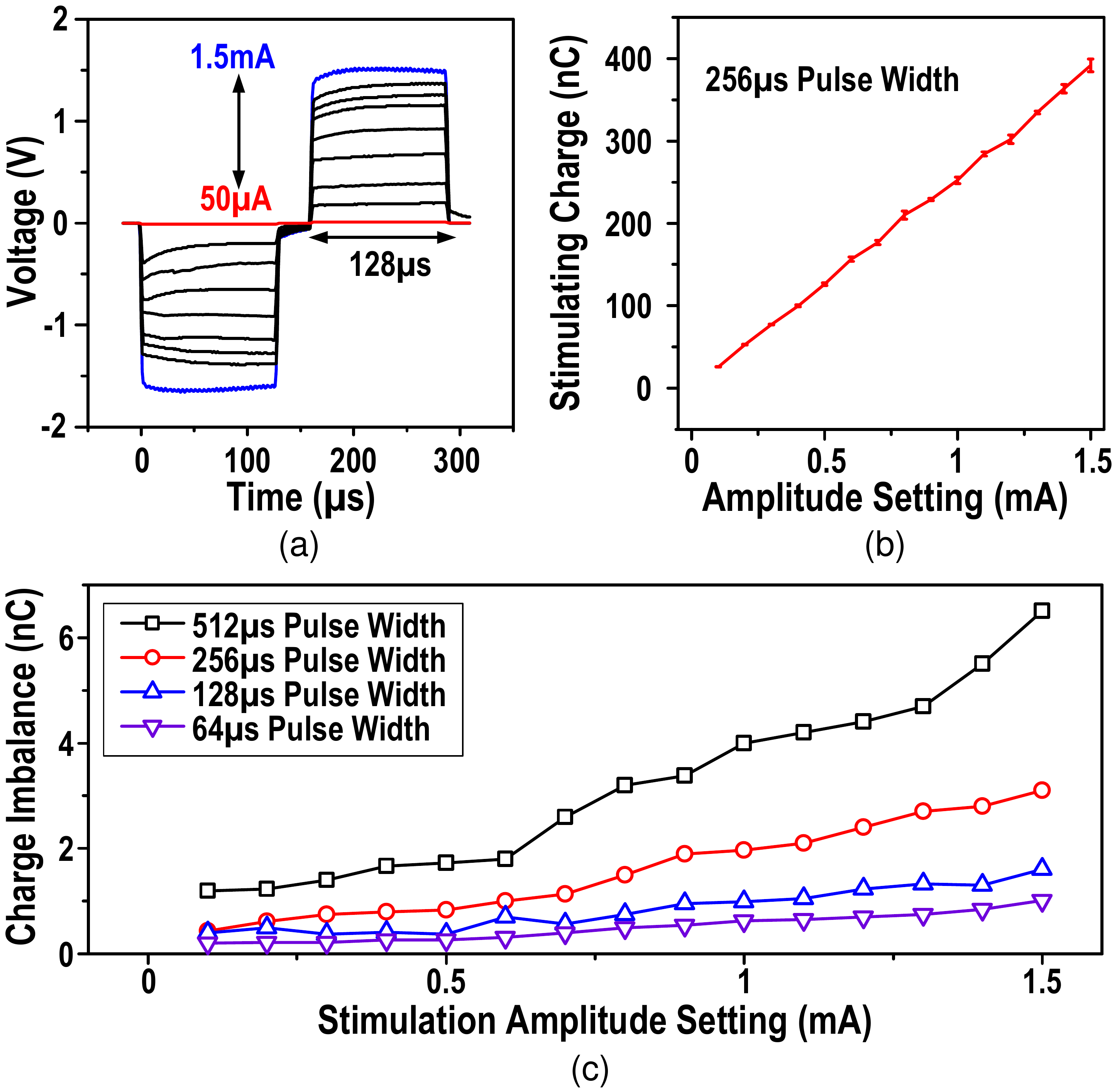}
      \caption{(a) Measured waveform of bi-phasic stimulation of 0.05-to-1.5-mA amplitudes and 256-\si{\micro}s pulse width, (b) stimulating charge at different amplitude settings with error bars indicating the variability and (c) charge imbalance of various stimulation parameters.}
      \label{S4_Stim_Program}
   \end{figure}

The SoC of the implant is fabricated in 180-nm CMOS technology, as shown in Fig.~\ref{S4_Die}. To validate the functionalities of the proposed neural stimulating system, the device is tested with the ME film in air, and the AC magnetic field is generated by a portable magnetic field driver (Fig.~\ref{S4_TX}).

Fig.~\ref{S4_Waveform} shows the measured waveforms of MagNI operating at different stimulation frequencies. Since transitions between operating phases are completely controlled by the TX, the frequency can be configured from 0 to 200~Hz with practically infinite resolution. As shown in Fig.~\ref{S4_Zoomin_2}, $V_\mathrm{rect}$ peaks at 2.95~V with an 83\% voltage conversion ratio at 0.61-mT magnetic flux density in the charging phase and drops to 2.55~V after the 1.5-mA, 512-\si{\micro}s stimulation. ASK modulation, which is actually a frequency modulation at TX, is realized by slightly shifting the magnetic field frequency, which causes the ME film to vibrate off resonance with reduced amplitude. Considering the settling time of the ME film and the fact that high data rate is not demanded in this application, 32 cycles are required to reliably transmit one bit, resulting in a 7.8-kbps data rate. 

Additionally, the waveform of bi-phasic stimulation current with various amplitudes is given in Fig.~\ref{S4_Stim_Program}(a). Total stimulation charge at different settings with error bars indicating the variability, whose maximum is 2\% of 1.5-mA amplitude, are shown in Fig.~\ref{S4_Stim_Program}(b). While the variability of stimulation caused by supply voltage variations and current source settling is observed, the error of the total charge deposited to the neural tissue is negligible. Because stimulation effects are believed to mostly depend on the total charge deposited to the neural tissue~\cite{cogan_neural_2008}, we believe the temporal stimulation amplitude variation will not affect the stimulation efficacy.
The asymmetry of bi-phasic stimulation may introduce charge imbalance. Total charge of the measured bi-phasic stimulation with different amplitudes and pulse widths is calculated, which shows the worst charge imbalance of 6.5 nC when the stimulation amplitude is 1.5~mA, 512~\si{\micro}s (Fig.~\ref{S4_Stim_Program}(c)). However, it should be noted that the electrodes are shorted after each stimulation, which removes the residual charge on the tissue and thus ensures the stimulation safety.


\subsection{In-Vitro Experiments}
\label{subsec:F_V_SoC}

\begin{figure}[t!]
      \centering
      \includegraphics[scale=0.6]{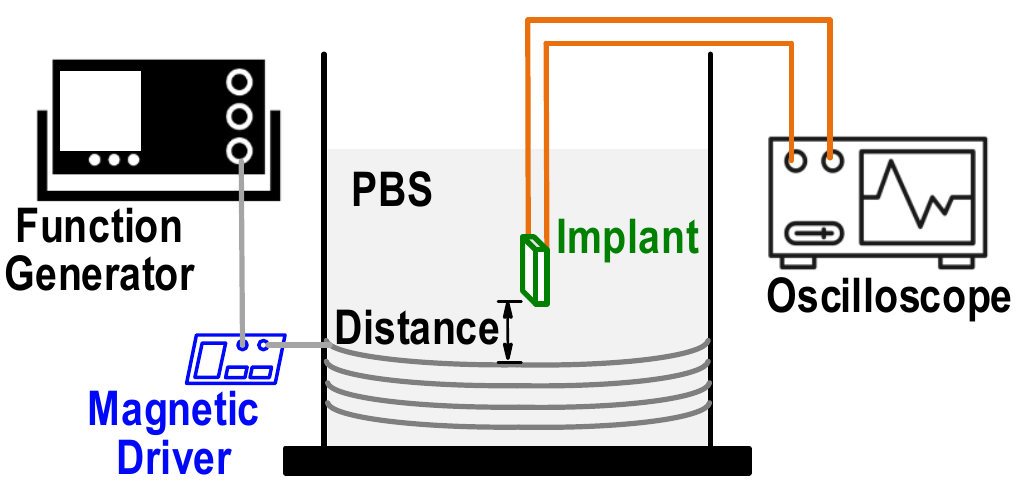}
      \caption{Diagram of the \textit{in-vitro} saline test setup with ME powering and access to implant for detailed operation analysis.}
      \label{S4_PBS_Setup}
   \end{figure}
   
\begin{figure}[t]
      \centering
      \includegraphics[scale=0.28]{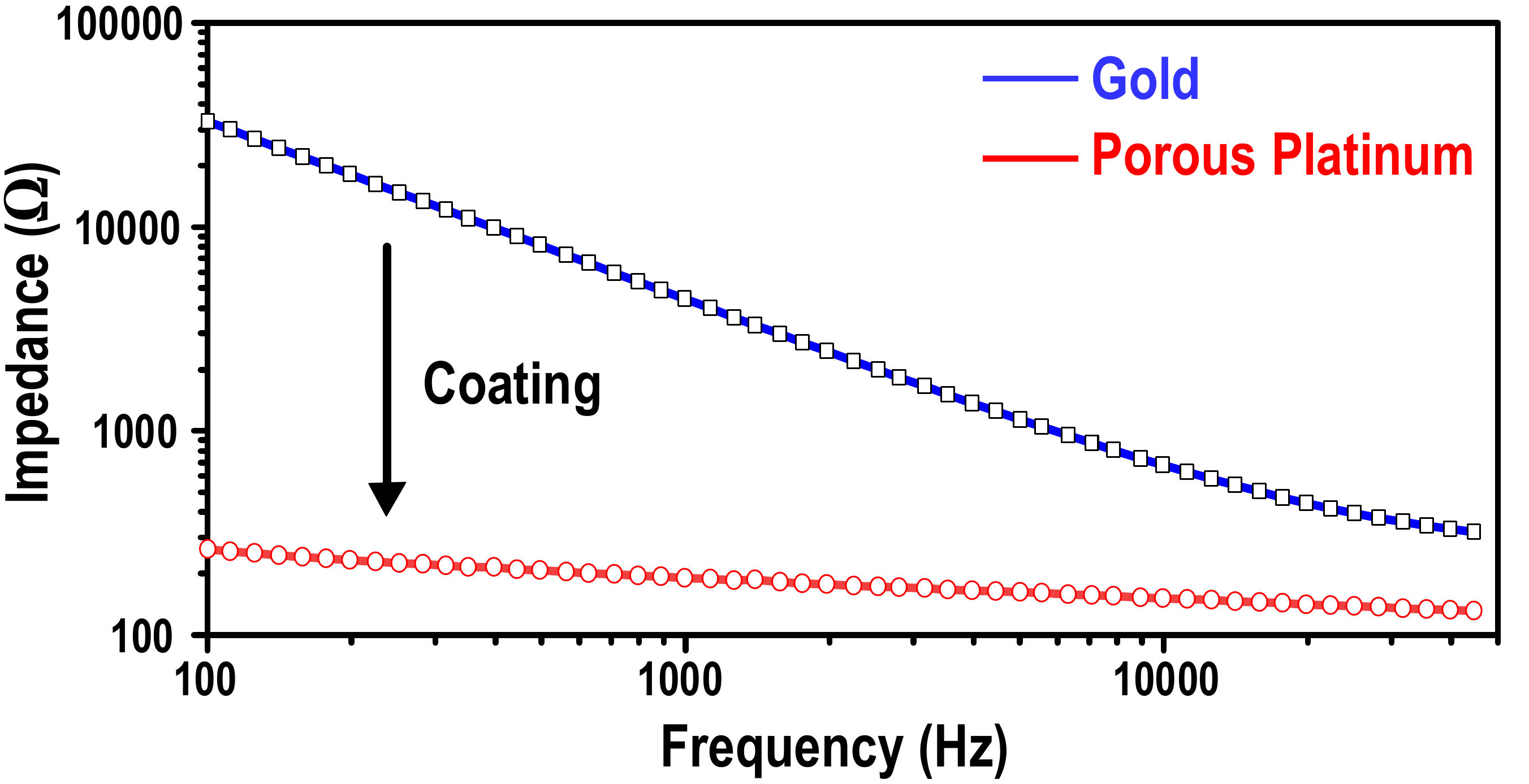}
      \caption{Impedance measurements of the on-board electrodes before and after porous platinum coating in PBS.}
      \label{S4_Elec}
   \end{figure}
   
\begin{figure}[t]
      \centering
      \includegraphics[scale=0.55]{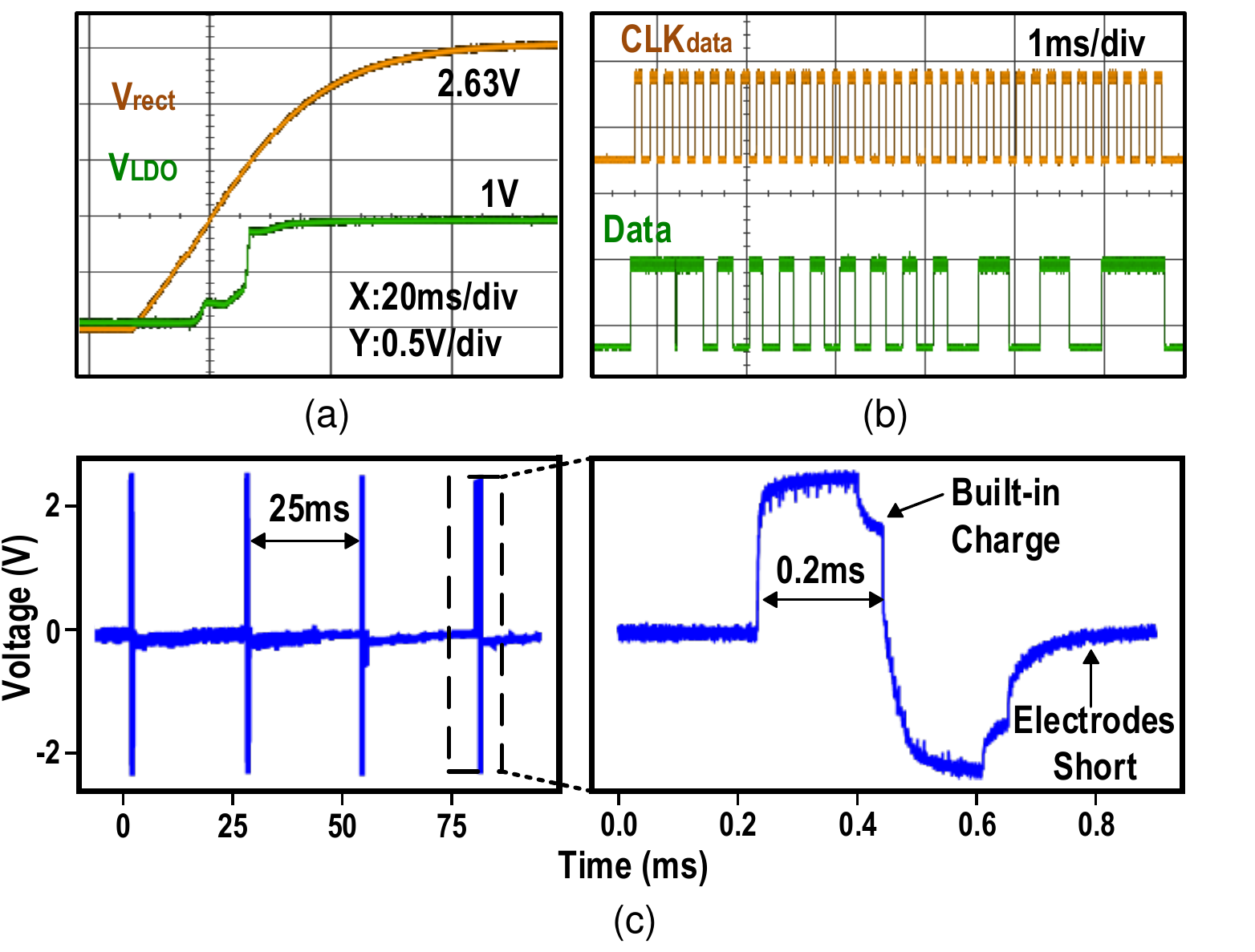}
      \caption{Measured (a) device startup, (b) data transfer and (c) stimulus in PBS.}
      \label{S4_PBS_Wave}
   \end{figure}

To demonstrate the compatibility for implantation, a soak test is performed by immersing a MagNI in phosphate buffered saline (PBS), as illustrated in Fig. \ref{S4_PBS_Setup}. The ME film is packaged in a 3D-printed 0.4-mm thick enclosure to eliminate environmental influences on its mechanical vibration before the entire implant is encapsulated with non-conductive epoxy. 

The on-board electrodes are tested in PBS to evaluate the voltage compliance of the device. Fig~\ref{S4_Elec} shows that the measured impedance of the bare electrodes with gold surface is 2100~\si{\Omega} at 2~kHz (associated with the 500-\si{\micro}s stimulation). After a porous platinum coating, the impedance drops to 170~\si{\Omega} \cite{fan_sputtered_2020}. Thus, the stimulation supply voltage, which is up to 3.3~V, is abundant to support the maximum amplitude of 1.5~mA. The order-of-magnitude impedance reduction by electrodes coating eliminates the needs of high-voltage circuits and improves the stimulation efficiency.

\begin{table}[t]
\caption{\textbf{Power Delivery versus Various TX-Implant Distances in PBS.}}
\label{table}
\centering
\setlength{\tabcolsep}{3.6pt}
\renewcommand{\arraystretch}{1.5}
\begin{tabular}{|p{40pt}|p{25pt}|p{50pt}|p{25pt}|p{25pt}|p{25pt}|}
\hline

\centering\arraybackslash {Distance (mm)} &
\centering\arraybackslash {$P_\mathrm{driver}$ (W)} &
\centering\arraybackslash {Magnetic Flux Density (mT)} & 
\centering\arraybackslash {$V_\mathrm{rect}$ (V)} & 
\centering\arraybackslash {$P_\mathrm{in,max}$ (mW)} &
\centering\arraybackslash {$\si{\eta}_\mathrm{max}$ (\%)}
\\\hline

\multicolumn{1}{|c|}{0}
& \multicolumn{1}{c|}{0.51} 
& \multicolumn{1}{c|}{0.62} 
& \multicolumn{1}{c|}{2.63}
& \multicolumn{1}{c|}{2.22} 
& \multicolumn{1}{c|}{0.435}
\\\hline

\multicolumn{1}{|c|}{7.5}
& \multicolumn{1}{c|}{1.2} 
& \multicolumn{1}{c|}{0.66} 
& \multicolumn{1}{c|}{2.61}
& \multicolumn{1}{c|}{2.16} 
& \multicolumn{1}{c|}{0.180}
\\\hline

\multicolumn{1}{|c|}{20}
& \multicolumn{1}{c|}{1.4} 
& \multicolumn{1}{c|}{0.48} 
& \multicolumn{1}{c|}{2.31}
& \multicolumn{1}{c|}{1.67}
& \multicolumn{1}{c|}{0.119}
\\\hline

\multicolumn{1}{|c|}{30}
& \multicolumn{1}{c|}{2.1} 
& \multicolumn{1}{c|}{0.13} 
& \multicolumn{1}{c|}{2.05}
& \multicolumn{1}{c|}{1.35} 
& \multicolumn{1}{c|}{0.064}
\\\hline

\end{tabular}
\label{tab1}
\end{table}

\begin{figure}[t!]
      \centering
      \includegraphics[scale=0.34]{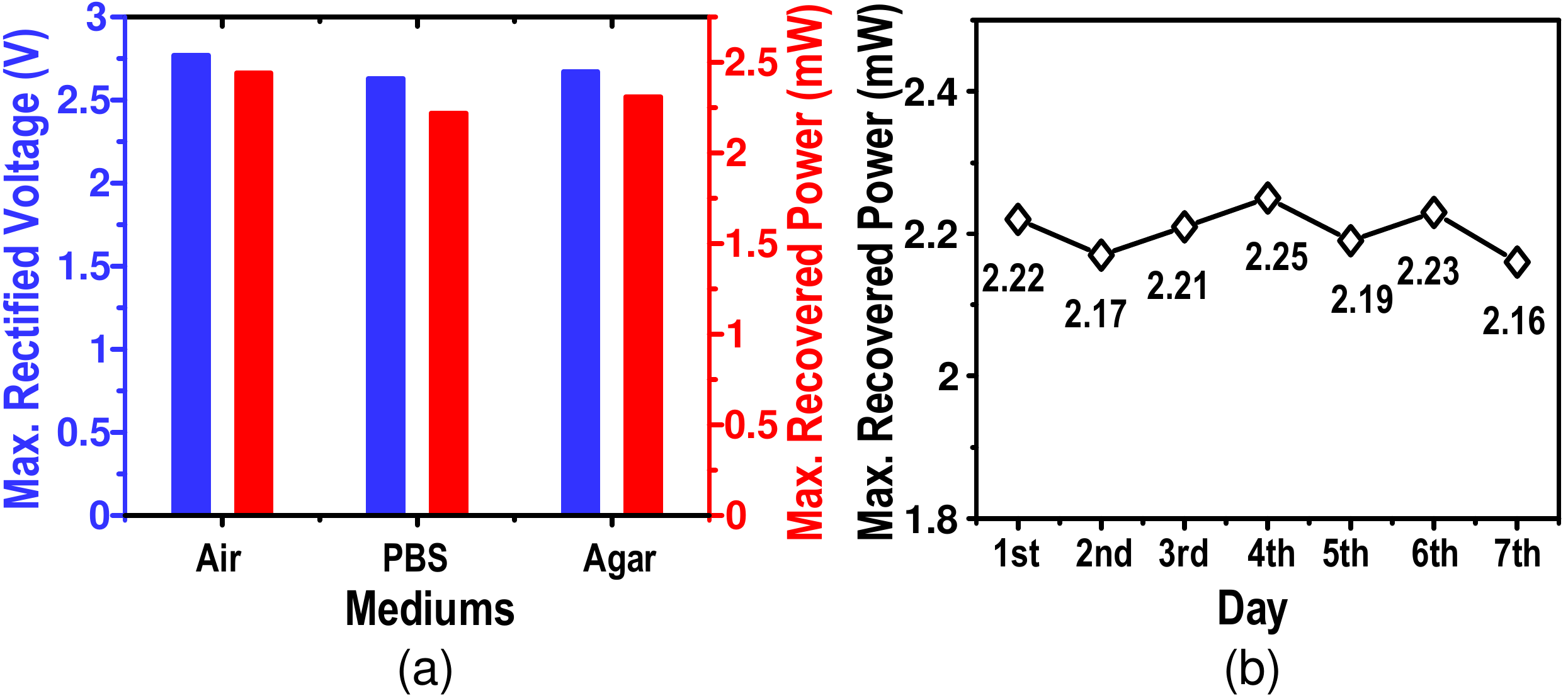}
      \caption{Measurements of (a) maximum recovered voltage and power in different mediums and (b) stability of power recovery throughout a 7-day soak test at the center of the TX coil.}
      \label{S4_Soak}
   \end{figure} 

During start-up (Fig.~\ref{S4_PBS_Wave}(a)), the device is charged up to 2.6~V in 44~ms with 2.22-mW peak harvested power $P_\mathrm{in,max}$ and 0.435-\% maximum power transfer efficiency , when at the center of the TX coil (Table~\ref{tab1}). 
Considering the 1-V source amplitude tolerance of the proposed implant, voltage drop with distance increase is allowed here to slow down the increment of TX power consumption and improve the overall efficiency. At a distance of 30~mm, $V_\mathrm{rect}$ of 2.05~V and $P_\mathrm{in,max}$ of 1.35~mW are achieved, which are sufficient to ensure SoC functionality.
$P_\mathrm{in,max}$ is estimated based on the measured rectified voltage $V_\mathrm{rect}$, as given by 
\begin{equation}
\begin{split}
P_\mathrm{in,max} \
\si{\approx}\ \mathrm{max} \ (C_\mathrm{store} \frac{d V_\mathrm{rect}}{dt} \times V_\mathrm{rect}).
\end{split}
\label{Pinmax}
\end{equation}
Fig.~\ref{S4_PBS_Wave}(c) shows the measured stimulation in saline with uncoated electrodes, which is 1.5 mA and 400~\si{\micro}s causing a voltage crossing electrodes of 2.53 V. Cathodic pulses compensate the built-in charge and residual charge is removed by shorting electrodes after stimulation. Effective charging-balance is realized by the bi-phasic stimulation shape and electrodes shorting phase.

In addition to testing in PBS, the device is also tested in air and agar, a substrate used to emulate the brain’s mechanical properties, showing $V_\mathrm{rect}$ and $P_\mathrm{in,max}$ variations less than 0.14~V and 0.22~mW respectively when at the center of the TX coil (Fig.~\ref{S4_Soak}(a)), demonstrating the ME effect’s adaptability in different mediums. Throughout the one-week soak test, the device in PBS functions
consistently with $P_\mathrm{in,max}$ fluctuating between 2.16-to-2.25~mW, indicating its long-term reliability for implantation (Fig.~\ref{S4_Soak}(b)).

\begin{figure}[t!]
      \centering
      \includegraphics[scale=0.55]{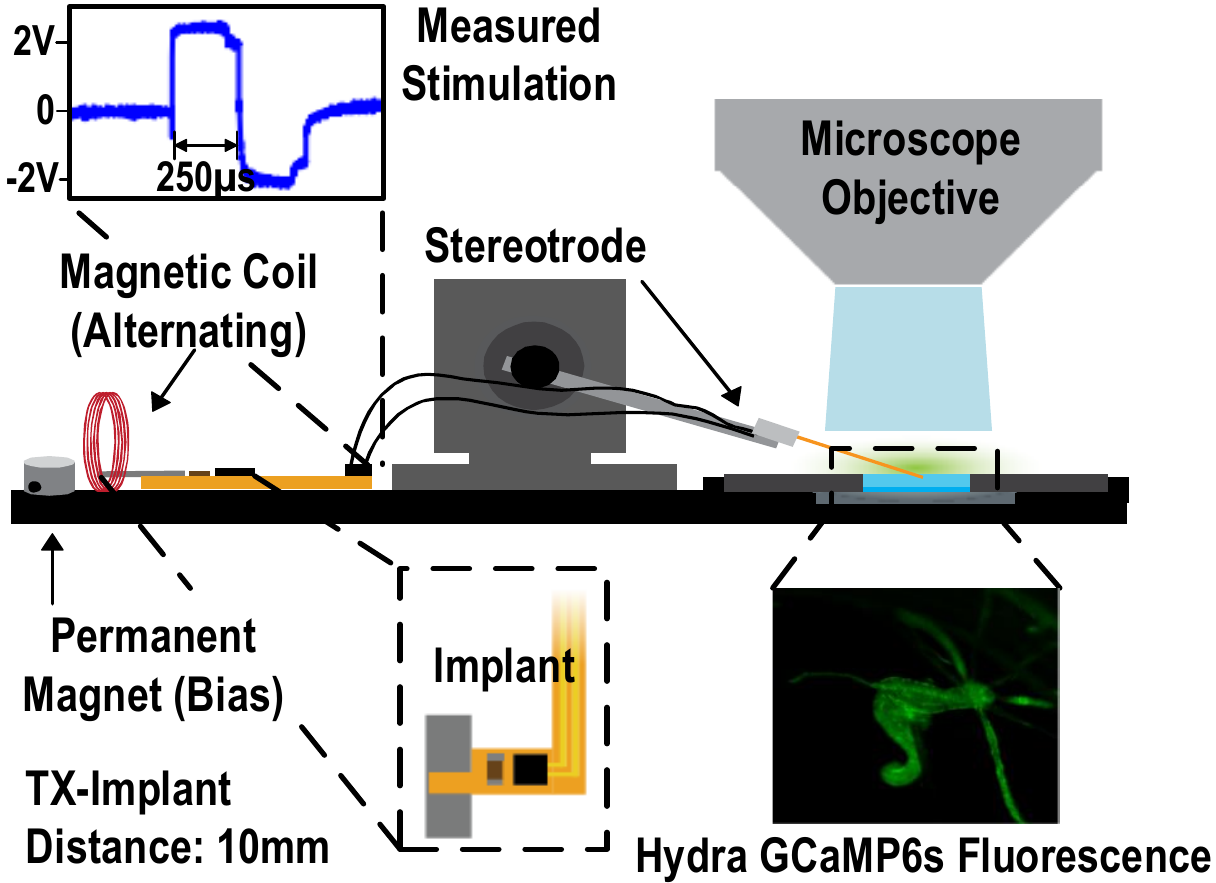}
      \caption{Diagram of the hydra stimulation test setup.}
      \label{S4_Hydra_Setup}
   \end{figure}

\begin{figure}[t!]
      \centering
      \includegraphics[scale=0.25]{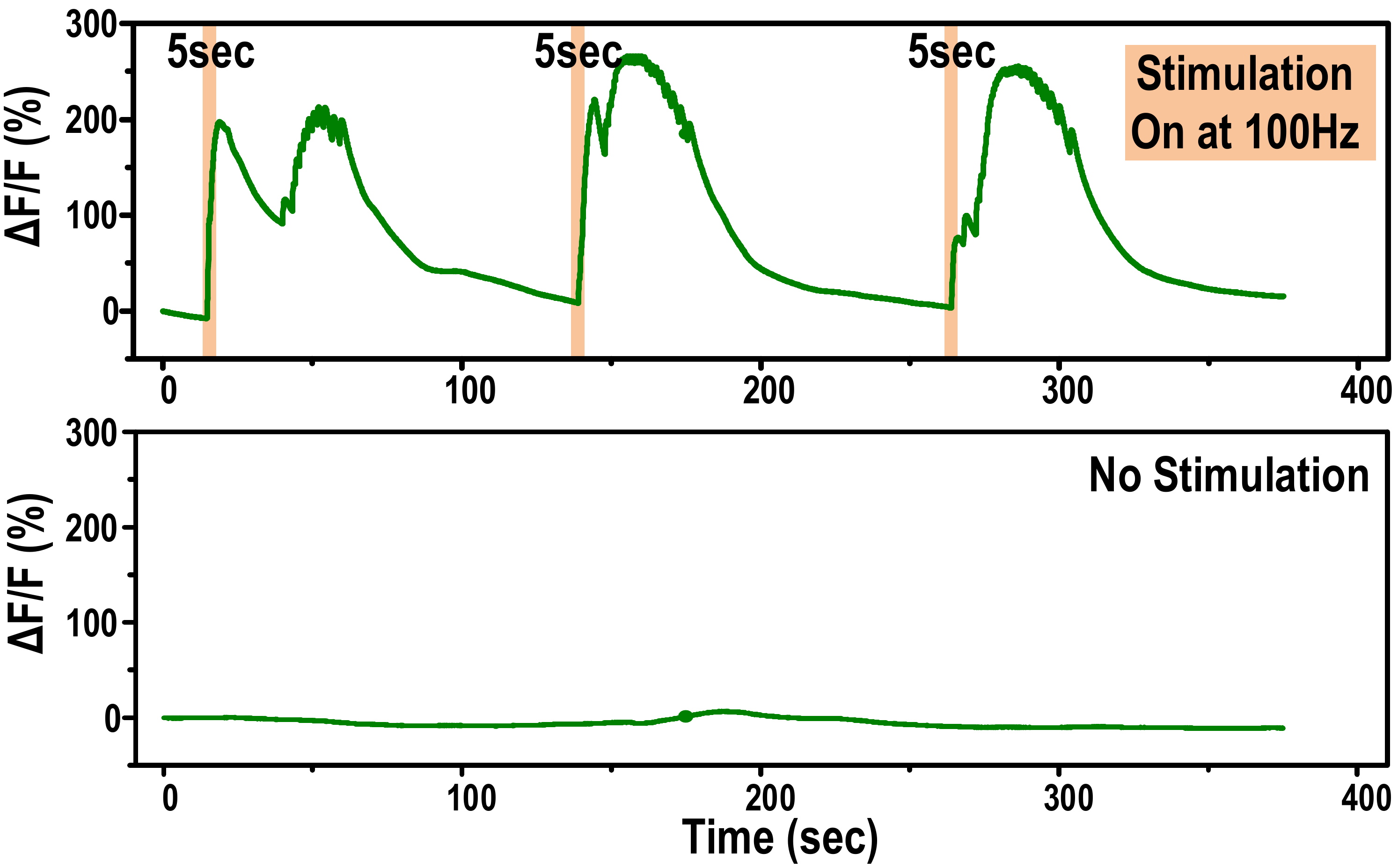}
      \caption{Time aligned measured fluorescence increases in response to electrical stimulations.}
      \label{S4_Hydra}
   \end{figure}  

\subsection{Hydra Stimulation Tests}
\label{subsec:Hydra}
   
To further assess MagNI’s bio-stimulation capability, the device is validated using \textit{Hydra vulgaris} as a model for excitable tissue~\cite{wittlieb_transgenic_2006}\cite{szymanski_mapping_2019}. The transgenic Hydra strains used express GCaMP6s, a calcium-sensitive fluorescent protein, in the ectoderm, and naturally express voltage sensitive ion channels. Fig. \ref{S4_Hydra_Setup} gives the diagram of the experimental setup using an inverted Nikon microscope with a 20x objective (NA = 0.75) and a constant excitation light of 460~nm. Fluorescence images are captured at 25 frames per second with 40~ms exposures and 2 x 2 binning using an Andor Zyla
sCMOS camera. In order to facilitate simultaneous stimulation and fluorescent imaging of Hydra, stereotrodes are used as the stimulating electrodes as they are proportional to the size of our chosen animal model and can more easily provide targeted stimulation to the area of interest. 
A micromanipulator is used to position the stereotrode connected to the proposed device that is remotely powered by the ME effect (10 mm from the TX); a blue laser is used to image GCaMP6s activity in Hydra. 

When we apply 5-sec biphasic pulse trains with 500-\si{\micro}s pulse widths at 100~Hz, we observe an increase larger than 200\% in GCaMP6s fluorescence, which is consistent with electrical activation of voltage gated ion channels in Hydra tissue that result in muscle contractions, as shown in Fig. \ref{S4_Hydra}. 




   
\begin{table*}[t]
\caption{\textbf{Comparison table with State-of-the-Art mm-Scale Wireless Neurostimulators}}
\label{table}
\centering
\setlength{\tabcolsep}{3.6pt}
\renewcommand{\arraystretch}{1.5}
\begin{tabular}{|p{100pt}|p{62pt}|p{56pt}|p{56pt}|p{56pt}|p{56pt}|p{56pt}|}
\hline

& 
\parbox[c][0.8cm]{60pt}{\centering
{\textbf{This Work}}}&
\parbox[c][0.8cm]{56pt} {\centering{Neuron$'$17 \\ \cite{shin_flexible_2017}}}&
\parbox[c][0.8cm]{56pt} {\centering{ISSCC$'$18 \\ \cite{jia_mm-sized_2018}}}&
\parbox[c][0.8cm]{56pt}{\centering
{TBioCAS$'$18 \\ \cite{lyu_energy-efficient_2018}}}&
\parbox[c][0.8cm]{56pt} {\centering{TBioCAS$'$19 \\ \cite{khalifa_microbead_2019}}} &
\parbox[c][0.8cm]{56pt} {\centering{Nat. Biomed. \\ Eng.$'$20  \cite{piech_wireless_2020}}} 
\\\hline

\multicolumn{1}{|c|}{\textbf{Application Target}} & 
\multicolumn{1}{c|}{\textbf{Spinal Cord}}& 
\multicolumn{1}{c|}{Brain}& 
\multicolumn{1}{c|}{Brain}& 
\multicolumn{1}{c|}{PNS}&
\multicolumn{1}{c|}{Brain}&
\multicolumn{1}{c|}{PNS}
\\\hline

\multicolumn{1}{|c|}{\textbf{Process (nm)}} & 
\multicolumn{1}{c|}{\textbf{180}}& 
\multicolumn{1}{c|}{Discrete}&
\multicolumn{1}{c|}{350}& 
\multicolumn{1}{c|}{180}&
\multicolumn{1}{c|}{130}&
\multicolumn{1}{c|}{65}
\\\hline

\multicolumn{1}{|c|}{\textbf{Wireless Link}} 
& \multicolumn{1}{c|}{\textbf{Magnetoelectric}}
& \multicolumn{1}{c|}{Inductive} 
& \multicolumn{1}{c|}{Inductive}
& \multicolumn{1}{c|}{Inductive} 
& \multicolumn{1}{c|}{Inductive}
& \multicolumn{1}{c|}{Ultrasonic}
\\\hline

\multicolumn{1}{|c|}{\textbf{Carrier Freq. (MHz)}} 
& \multicolumn{1}{c|}{\textbf{0.25}}
& \multicolumn{1}{c|}{13.56}
& \multicolumn{1}{c|}{60}
& \multicolumn{1}{c|}{198}
& \multicolumn{1}{c|}{1180}
& \multicolumn{1}{c|}{1.85}
\\\hline

\multicolumn{1}{|c|}{\textbf{Implant Volume (mm$^3$)}}
& \multicolumn{1}{c|}{\textbf{8.2}}
& \multicolumn{1}{c|}{98}
& \multicolumn{1}{c|}{12.2}
& \multicolumn{1}{c|}{97.5}
& \multicolumn{1}{c|}{0.009}
& \multicolumn{1}{c|}{1.7}
\\\hline


\parbox[c][0.85cm]{100pt}{\centering\textbf
  {Stimulation Mode}}
& \parbox[c][0.8cm]{62pt}{\centering\textbf
  {Biphasic \\ Current}}
& \parbox[c][0.8cm]{56pt}{\centering
  {Optical}} 
& \parbox[c][0.8cm]{56pt}{\centering
  {Optical}}
& \parbox[c][0.8cm]{56pt}{\centering
  { Monophasic \\ Voltage}}
& \parbox[c][0.8cm]{56pt}{\centering
  {Monophasic \\ Voltage}}
& \parbox[c][0.8cm]{56pt}{\centering
  {Monophasic \\ Current}}
\\\hline

\parbox[c][0.85cm]{100pt}{\centering\textbf
  {Max. Stim. Current (mA) / \\
Resolution (bit)}}
& \multicolumn{1}{c|}{\textbf{1.5 / 5}}
& \multicolumn{1}{c|}{N/A}
& \multicolumn{1}{c|}{N/A}
& \multicolumn{1}{c|}{0.25 / Fixed}
& \multicolumn{1}{c|}{0.04 / Fixed}
& \multicolumn{1}{c|}{0.4 / 3}
\\\hline

\multicolumn{1}{|c|}{\textbf{Max. Stimulation Charge (nC) }}
& \multicolumn{1}{c|}{\textbf{391}}
& \multicolumn{1}{c|}{N/A}
& \multicolumn{1}{c|}{N/A}
& \multicolumn{1}{c|}{17.5}
& \multicolumn{1}{c|}{0.66}
& \multicolumn{1}{c|}{157}
\\\hline

\multicolumn{1}{|c|}{\textbf{Stimulating Channels}}
& \multicolumn{1}{c|}{\textbf{1}}
& \multicolumn{1}{c|}{1}
& \multicolumn{1}{c|}{16}
& \multicolumn{1}{c|}{1}
& \multicolumn{1}{c|}{1}
& \multicolumn{1}{c|}{1}
\\\hline

\multicolumn{1}{|c|}{\textbf{SoC Power (\si{\micro}W)}}  
& \multicolumn{1}{c|}{\textbf{23.7}} 
& \multicolumn{1}{c|}{N/A}
& \multicolumn{1}{c|}{300}
& \multicolumn{1}{c|}{2.7}
& \multicolumn{1}{c|}{N/A}
& \multicolumn{1}{c|}{4}
\\\hline


\parbox[c][0.85cm]{100pt}{\centering\textbf
  {Max. Overall Efficiency (\%) \\
@ Distance (mm)}}
& \parbox[c][0.8cm]{60pt}{\centering\textbf
  {0.435 @ 0 \\ 0.064 @ 30}}
& \multicolumn{1}{c|}{N/A}
& \multicolumn{1}{c|}{N/A}
& \multicolumn{1}{c|}{N/A}
& \multicolumn{1}{c|}{0.0019 @ 6.6}
& \multicolumn{1}{c|}{0.06 @ 18}
\\\hline


\multicolumn{1}{|c|}{\textbf{Max. Operating Depth (mm)}}
& \multicolumn{1}{c|}{\textbf{30}}
& \multicolumn{1}{c|}{40}
& \multicolumn{1}{c|}{7}
& \multicolumn{1}{c|}{140}
& \multicolumn{1}{c|}{6.6}
& \multicolumn{1}{c|}{55}
\\\hline


\multicolumn{1}{|c|}{\textbf{In-Vivo Test Model}}
& \multicolumn{1}{c|}{\textbf{Hydra}}
& \multicolumn{1}{c|}{Rat}
& \multicolumn{1}{c|}{Rat}
& \multicolumn{1}{c|}{Rat}
& \multicolumn{1}{c|}{Rat}
& \multicolumn{1}{c|}{Rat}
\\\hline

\end{tabular}
\label{tab2}
\end{table*}

\section{Conclusion}
This work presents a proof-of-concept demonstration of MagNI, a magnetoelectrically powered and controlled neural implant. ME power link proves to be a viable technology to wirelessly deliver greater than 1 mW power to mm-sized devices deeply implanted in the body at 30-mm depth. The proposed spinal cord stimulator for neuropathic pain relief features miniaturized physical dimensions of 8.2 mm$^3$ and 28 mg, adaptive operation and data transfer mechanisms (1-V source variation tolerance), and programmable bi-phasic current stimulation capability fully covering 0.05 to 1.5-mA amplitude, 64 to 512-\si{\micro}s pulse width, and 0 to 200-Hz repetition range. 
Table~\ref{table} summarizes the specifications of MagNI and compares with state-of-the-art mm-scale wireless neural stimulators. The use of low-frequency magnetic field  (250 kHz) substantially alleviates tissue absorption and reflections in wireless power transfer, leading to higher deliverable power under safety limits (in Section II). Thanks to the high power transfer efficiency and chip efficiency, the proposed system achieves one of the highest 391-nC total stimulation charges among the comparisons, with 0.435\% and 0.064\% end-to-end efficiency at 0 and 30mm distance respectively. In addition, ME power link features robustness against alignment changes (20\% less power with 50 degree angular mismatch between TX and implants).


%

\appendices


\section*{Acknowledgment}

The authors would like to thank Christophe Dupre and Dr. Rafael Yuste for sharing the transgenic Hydra strains. We would also like to thank Soonyoung Kim for the culturing and handling of Hydra and useful discussions of Hydra biology.

\ifCLASSOPTIONcaptionsoff
  \newpage
\fi



\bibliography{bibtex/bib/Background.bib, bibtex/bib/Optical.bib, bibtex/bib/Ultrasound.bib, bibtex/bib/Inductive.bib, bibtex/bib/RF.bib, bibtex/bib/ME.bib, bibtex/bib/Tissue.bib, bibtex/bib/Circuit.bib,
bibtex/bib/Revision.bib}

\begin{thebibliography}{10}

\bibitem{shealy_electrical_1967}
C.~N. Shealy, J.~T. Mortimer, and J.~B. Reswick, ``Electrical {Inhibition} of
  {Pain} by {Stimulation} of the {Dorsal} {Columns}: {Preliminary} {Clinical}
  {Report},'' {\em Anesthesia \& Analgesia}, vol.~46, pp.~489--491, July 1967.

\bibitem{verrills_review_2016}
P.~Verrills, C.~Sinclair, and A.~Barnard, ``A review of spinal cord stimulation
  systems for chronic pain,'' {\em J Pain Res}, vol.~9, pp.~481--492, July
  2016.

\bibitem{hargreaves_kirschner_2004}
D.~G. Hargreaves, S.~J. Drew, and R.~Eckersley, ``Kirschner {Wire} {Pin}
  {Tract} {Infection} {Rates}: {A} randomized controlled trial between
  percutaneous and buried wires,'' {\em The Journal of Hand Surgery: British \&
  European Volume}, vol.~29, pp.~374--376, Aug. 2004.

\bibitem{biran_brain_2007}
R.~Biran, D.~C. Martin, and P.~A. Tresco, ``The brain tissue response to
  implanted silicon microelectrode arrays is increased when the device is
  tethered to the skull,'' {\em Journal of Biomedical Materials Research Part
  A}, vol.~82A, no.~1, pp.~169--178, 2007.

\bibitem{pinnell_miniaturized_2018}
R.~C. Pinnell, A.~Pereira~de Vasconcelos, J.~C. Cassel, and U.~G. Hofmann, ``A
  {Miniaturized}, {Programmable} {Deep}-{Brain} {Stimulator} for
  {Group}-{Housing} and {Water} {Maze} {Use},'' {\em Front. Neurosci.},
  vol.~12, 2018.

\bibitem{montgomery_wirelessly_2015}
K.~L. Montgomery, A.~J. Yeh, {\em et~al.}, ``Wirelessly powered, fully internal
  optogenetics for brain, spinal and peripheral circuits in mice,'' {\em Nature
  Methods}, vol.~12, pp.~969--974, Oct. 2015.

\bibitem{leung_cmos_2018}
V.~W. Leung, J.~Lee, S.~Li, {\em et~al.}, ``A {CMOS} {Distributed} {Sensor}
  {System} for {High}-{Density} {Wireless} {Neural} {Implants} for
  {Brain}-{Machine} {Interfaces},'' in {\em {ESSCIRC} 2018 - {IEEE} 44th
  {European} {Solid} {State} {Circuits} {Conference} ({ESSCIRC})},
  pp.~230--233, Sept. 2018.

\bibitem{biederman_fully-integrated_2013}
W.~Biederman, D.~J. Yeager, N.~Narevsky, {\em et~al.}, ``A
  {Fully}-{Integrated}, {Miniaturized} 0.125 mm$^2$ 10.5 \si{\micro}{W}
  {Wireless} {Neural} {Sensor},'' {\em IEEE Journal of Solid-State Circuits},
  vol.~48, pp.~960--970, Apr. 2013.

\bibitem{lee_power-efficient_2015}
H.-M. Lee, K.~Y. Kwon, W.~Li, and M.~Ghovanloo, ``A {Power}-{Efficient}
  {Switched}-{Capacitor} {Stimulating} {System} for {Electrical}/{Optical}
  {Deep} {Brain} {Stimulation},'' {\em IEEE Journal of Solid-State Circuits},
  vol.~50, pp.~360--374, Jan. 2015.

\bibitem{lo_176-channel_2016}
Y.-K. Lo, C.-W. Chang, Y.-C. Kuan, {\em et~al.}, ``A 176-channel 0.5 cm$^3$ 0.7
  g wireless implant for motor function recovery after spinal cord injury,'' in
  {\em 2016 {IEEE} {International} {Solid}-{State} {Circuits} {Conference}
  ({ISSCC})}, pp.~382--383, Jan. 2016.

\bibitem{shin_flexible_2017}
G.~Shin, A.~M. Gomez, R.~Al-Hasani, {\em et~al.}, ``Flexible {Near}-{Field}
  {Wireless} {Optoelectronics} as {Subdermal} {Implants} for {Broad}
  {Applications} in {Optogenetics},'' {\em Neuron}, vol.~93, pp.~509--521.e3,
  Feb. 2017.

\bibitem{freeman_sub-millimeter_2017}
D.~K. Freeman, J.~M. O'Brien, P.~Kumar, {\em et~al.}, ``A {Sub}-millimeter,
  {Inductively} {Powered} {Neural} {Stimulator},'' {\em Front. Neurosci.},
  vol.~11, 2017.

\bibitem{jia_mm-sized_2018}
Y.~Jia, S.~A. Mirbozorgi, B.~Lee, {\em et~al.}, ``A mm-sized free-floating
  wirelessly powered implantable optical stimulating system-on-a-chip,'' in
  {\em 2018 {IEEE} {International} {Solid} - {State} {Circuits} {Conference} -
  ({ISSCC})}, pp.~468--470, Feb. 2018.

\bibitem{lyu_energy-efficient_2018}
H.~Lyu, J.~Wang, J.-H. La, J.~M. Chung, and A.~Babakhani, ``An
  {Energy}-{Efficient} {Wirelessly} {Powered} {Millimeter}-{Scale}
  {Neurostimulator} {Implant} {Based} on {Systematic} {Codesign} of an
  {Inductive} {Loop} {Antenna} and a {Custom} {Rectifier},'' {\em IEEE
  Transactions on Biomedical Circuits and Systems}, vol.~12, pp.~1131--1143,
  Oct. 2018.

\bibitem{khalifa_microbead_2019}
A.~Khalifa, Y.~Liu, Y.~Karimi, {\em et~al.}, ``The {Microbead}: {A} 0.009
  mm$^3$ {Implantable} {Wireless} {Neural} {Stimulator},'' {\em IEEE
  Transactions on Biomedical Circuits and Systems}, vol.~13, pp.~971--985, Oct.
  2019.

\bibitem{burton_wireless_2020}
A.~Burton, S.~N. Obaid, A.~Vázquez-Guardado, {\em et~al.}, ``Wireless,
  battery-free subdermally implantable photometry systems for chronic recording
  of neural dynamics,'' {\em Proceedings of the National Academy of Sciences},
  vol.~117, pp.~2835--2845, Feb. 2020.

\bibitem{charthad_mm-sized_2015}
J.~Charthad, M.~J. Weber, T.~C. Chang, and A.~Arbabian, ``A mm-{Sized}
  {Implantable} {Medical} {Device} ({IMD}) {With} {Ultrasonic} {Power}
  {Transfer} and a {Hybrid} {Bi}-{Directional} {Data} {Link},'' {\em IEEE
  Journal of Solid-State Circuits}, vol.~50, pp.~1741--1753, Aug. 2015.

\bibitem{seo_wireless_2016}
D.~Seo, R.~M. Neely, K.~Shen, {\em et~al.}, ``Wireless {Recording} in the
  {Peripheral} {Nervous} {System} with {Ultrasonic} {Neural} {Dust},'' {\em
  Neuron}, vol.~91, pp.~529--539, Aug. 2016.

\bibitem{charthad_mm-sized_2018}
J.~Charthad, T.~C. Chang, Z.~Liu, {\em et~al.}, ``A mm-{Sized} {Wireless}
  {Implantable} {Device} for {Electrical} {Stimulation} of {Peripheral}
  {Nerves},'' {\em IEEE Transactions on Biomedical Circuits and Systems},
  vol.~12, pp.~257--270, Apr. 2018.

\bibitem{ghanbari_sub-mm3_2019}
M.~M. Ghanbari, D.~K. Piech, K.~Shen, {\em et~al.}, ``A {Sub}-mm3 {Ultrasonic}
  {Free}-{Floating} {Implant} for {Multi}-{Mote} {Neural} {Recording},'' {\em
  IEEE Journal of Solid-State Circuits}, vol.~54, pp.~3017--3030, Nov. 2019.

\bibitem{piech_wireless_2020}
D.~K. Piech, B.~C. Johnson, K.~Shen, {\em et~al.}, ``A wireless
  millimetre-scale implantable neural stimulator with ultrasonically powered
  bidirectional communication,'' {\em Nature Biomedical Engineering}, vol.~4,
  pp.~207--222, Feb. 2020.

\bibitem{lee_250_2018}
S.~Lee, A.~J. Cortese, A.~P. Gandhi, {\em et~al.}, ``A 250 \si{\micro}m × 57
  \si{\micro}m {Microscale} {Opto}-electronically {Transduced} {Electrodes}
  ({MOTEs}) for {Neural} {Recording},'' {\em IEEE Transactions on Biomedical
  Circuits and Systems}, vol.~12, pp.~1256--1266, Dec. 2018.

\bibitem{lim_019017mm2_2020}
J.~Lim, E.~Moon, M.~Barrow, {\em et~al.}, ``A 0.19 × 0.17 mm$^2$ {Wireless}
  {Neural} {Recording} {IC} for {Motor} {Prediction} with
  {Near}-{Infrared}-{Based} {Power} and {Data} {Telemetry},'' in {\em 2020
  {IEEE} {International} {Solid}- {State} {Circuits} {Conference} - ({ISSCC})},
  pp.~416--418, Feb. 2020.

\bibitem{noauthor_ieee_nodate}
``{IEEE} {Standard} for {Safety} {Levels} with {Respect} to {Human} {Exposure}
  to {Electric}, {Magnetic}, and {Electromagnetic} {Fields}, 0 {Hz} to 300
  {GHz},'' tech. rep., IEEE.
\newblock ISBN: 9781504455480.

\bibitem{denisov_ultrasonic_2010}
A.~Denisov and E.~Yeatman, ``Ultrasonic vs. {Inductive} {Power} {Delivery} for
  {Miniature} {Biomedical} {Implants},'' in {\em 2010 {International}
  {Conference} on {Body} {Sensor} {Networks}}, pp.~84--89, June 2010.

\bibitem{hoskins_diagnostic_2010}
P.~R. Hoskins, K.~Martin, and A.~Thrush, {\em Diagnostic {Ultrasound}:
  {Physics} and {Equipment}}.
\newblock Cambridge University Press, June 2010.

\bibitem{chen_wireless_2015}
R.~Chen, G.~Romero, {\em et~al.}, ``Wireless magnetothermal deep brain
  stimulation,'' {\em Science}, vol.~347, pp.~1477--1480, Mar. 2015.

\bibitem{zaeimbashi_nanoneurorfid_2019}
M.~Zaeimbashi, H.~Lin, C.~Dong, {\em et~al.}, ``{NanoNeuroRFID}: {A} {Wireless}
  {Implantable} {Device} {Based} on {Magnetoelectric} {Antennas},'' {\em IEEE
  Journal of Electromagnetics, RF and Microwaves in Medicine and Biology},
  vol.~3, pp.~206--215, Sept. 2019.

\bibitem{singer_magnetoelectric_2020}
A.~Singer, S.~Dutta, E.~Lewis, {\em et~al.}, ``Magnetoelectric {Materials} for
  {Miniature}, {Wireless} {Neural} {Stimulation} at {Therapeutic}
  {Frequencies},'' {\em Neuron}, June 2020.

\bibitem{yu_82mm3_2020}
Z.~Yu, J.~C. Chen, B.~W. Avants, {\em et~al.}, ``An 8.2 mm$^3$ {Implantable}
  {Neurostimulator} with {Magnetoelectric} {Power} and {Data} {Transfer},'' in
  {\em 2020 {IEEE} {International} {Solid}- {State} {Circuits} {Conference} -
  ({ISSCC})}, pp.~510--512, Feb. 2020.

\bibitem{shuxiang_dong_longitudinal_2003}
S.~Dong, J.-F. Li, and D.~Viehland, ``Longitudinal and transverse
  magnetoelectric voltage coefficients of magnetostrictive/piezoelectric
  laminate composite: theory,'' {\em IEEE Transactions on Ultrasonics,
  Ferroelectrics, and Frequency Control}, vol.~50, pp.~1253--1261, Oct. 2003.

\bibitem{zhai_giant_2006}
J.~Zhai, S.~Dong, Z.~Xing, J.~Li, and D.~Viehland, ``Giant magnetoelectric
  effect in {Metglas}/polyvinylidene-fluoride laminates,'' {\em Applied Physics
  Letters}, vol.~89, p.~083507, Aug. 2006.

\bibitem{silva_optimization_2013}
M.~Silva, S.~Reis, {\em et~al.}, ``Optimization of the {Magnetoelectric}
  {Response} of {Poly}(vinylidene fluoride)/{Epoxy}/{Vitrovac} {Laminates},''
  {\em ACS Applied Materials \& Interfaces}, vol.~5, pp.~10912--10919, Nov.
  2013.

\bibitem{dong_equivalent_2008}
S.~Dong and J.~Zhai, ``Equivalent circuit method for static and dynamic
  analysis of magnetoelectric laminated composites,'' {\em Chin. Sci. Bull.},
  vol.~53, pp.~2113--2123, July 2008.

\bibitem{zhou_uniform_2014}
J.-P. Zhou, Y.-J. Ma, G.-B. Zhang, and X.-M. Chen, ``A uniform model for direct
  and converse magnetoelectric effect in laminated composite,'' {\em Applied
  Physics Letters}, vol.~104, p.~202904, May 2014.

\bibitem{young_frequencydepth-penetration_1980}
J.~Young, M.-T. Wang, and I.~Brezovich, ``Frequency/depth-penetration
  considerations in hyperthermia by magnetically induced currents,'' {\em
  Electronics Letters}, vol.~16, pp.~358--359, May 1980.

\bibitem{hasgall_itis_2015}
P.~Hasgall, E.~Neufeld, M.-C. Gosselin, A.~Klingenböck, and N.~Kuster,
  ``{ITIS} {Database} for thermal and electromagnetic parameters of biological
  tissues, {Version} 4.0,'' May 2015.

\bibitem{shellock_implantable_2004}
F.~G. Shellock, G.~Cosendai, S.-M. Park, and J.~A. Nyenhuis, ``Implantable
  {Microstimulator}: {Magnetic} {Resonance} {Safety} at 1.5 {Tesla},'' {\em
  Investigative Radiology}, vol.~39, pp.~591--599, Oct. 2004.

\bibitem{fang_enhancing_2009}
Z.~Fang, S.~G. Lu, F.~Li, {\em et~al.}, ``Enhancing the magnetoelectric
  response of {Metglas}/polyvinylidene fluoride laminates by exploiting the
  flux concentration effect,'' {\em Applied Physics Letters}, vol.~95,
  p.~112903, Sept. 2009.

\bibitem{chen_245-ghz_2008}
X.~Chen, W.~G. Yeoh, Y.~B. Choi, H.~Li, and R.~Singh, ``A 2.45-{GHz}
  {Near}-{Field} {RFID} {System} {With} {Passive} {On}-{Chip} {Antenna}
  {Tags},'' {\em IEEE Transactions on Microwave Theory and Techniques},
  vol.~56, pp.~1397--1404, June 2008.

\bibitem{landgren_wideband_2017}
D.~W. Landgren, K.~R. Cook, D.~J.~P. Dykes, {\em et~al.}, ``A wideband {mmWave}
  antenna element with an unbalanced feed,'' in {\em 2017 {IEEE} {National}
  {Aerospace} and {Electronics} {Conference} ({NAECON})}, pp.~209--212, June
  2017.

\bibitem{truong_experimentally_2020}
B.~D. Truong and S.~Roundy, ``Experimentally validated model and power
  optimization of a magnetoelectric wireless power transfer system in free-free
  configuration,'' {\em Smart Materials and Structures}, May 2020.

\bibitem{kuo_equation-based_2018}
N.-C. Kuo, B.~Zhao, and A.~M. Niknejad, ``Equation-{Based} {Optimization} for
  {Inductive} {Power} {Transfer} to a {Miniature} {CMOS} {Rectenna},'' {\em
  IEEE Transactions on Microwave Theory and Techniques}, vol.~66,
  pp.~2393--2408, May 2018.

\bibitem{truong_fundamental_2020}
B.~D. Truong, ``Fundamental {Issues} in {Magnetoelectric} {Transducers}:
  {Magnetic} {Field} {Sensing} {Versus} {Wireless} {Power} {Transfer}
  {Systems},'' {\em IEEE Sensors Journal}, vol.~20, pp.~5322--5328, May 2020.

\bibitem{lam_integrated_2006}
Y.-H. Lam, W.-H. Ki, and C.-Y. Tsui, ``Integrated {Low}-{Loss} {CMOS} {Active}
  {Rectifier} for {Wirelessly} {Powered} {Devices},'' {\em IEEE Transactions on
  Circuits and Systems II: Express Briefs}, vol.~53, pp.~1378--1382, Dec. 2006.

\bibitem{seok_portable_2012}
M.~Seok, G.~Kim, {\em et~al.}, ``A {Portable} 2-{Transistor} {Picowatt}
  {Temperature}-{Compensated} {Voltage} {Reference} {Operating} at 0.5 {V},''
  {\em IEEE Journal of Solid-State Circuits}, vol.~47, pp.~2534--2545, Oct.
  2012.

\bibitem{gorman_effect_1983}
P.~H. Gorman and J.~T. Mortimer, ``The {Effect} of {Stimulus} {Parameters} on
  the {Recruitment} {Characteristics} of {Direct} {Nerve} {Stimulation},'' {\em
  IEEE Transactions on Biomedical Engineering}, vol.~BME-30, pp.~407--414, July
  1983.

\bibitem{cogan_neural_2008}
S.~F. Cogan, ``Neural {Stimulation} and {Recording} {Electrodes},'' {\em Annual
  Review of Biomedical Engineering}, vol.~10, no.~1, pp.~275--309, 2008.

\bibitem{fan_sputtered_2020}
B.~Fan, A.~V. Rodriguez, D.~G. Vercosa, {\em et~al.}, ``Sputtered porous {Pt}
  for wafer-scale manufacture of low-impedance flexible microelectrodes,'' {\em
  Journal of Neural Engineering}, vol.~17, p.~036029, June 2020.

\bibitem{wittlieb_transgenic_2006}
J.~Wittlieb, K.~Khalturin, {\em et~al.}, ``Transgenic {Hydra} allow in vivo
  tracking of individual stem cells during morphogenesis,'' {\em Proceedings of
  the National Academy of Sciences}, vol.~103, pp.~6208--6211, Apr. 2006.

\bibitem{szymanski_mapping_2019}
J.~R. Szymanski and R.~Yuste, ``Mapping the {Whole}-{Body} {Muscle} {Activity}
  of {Hydra} vulgaris,'' {\em Current Biology}, vol.~29, pp.~1807--1817.e3,
  June 2019.

\end{thebibliography}
\bibliographystyle{ieeetr}
%



%

\begin{IEEEbiography}[{\includegraphics[width=1in,height=1.25in,clip,keepaspectratio]{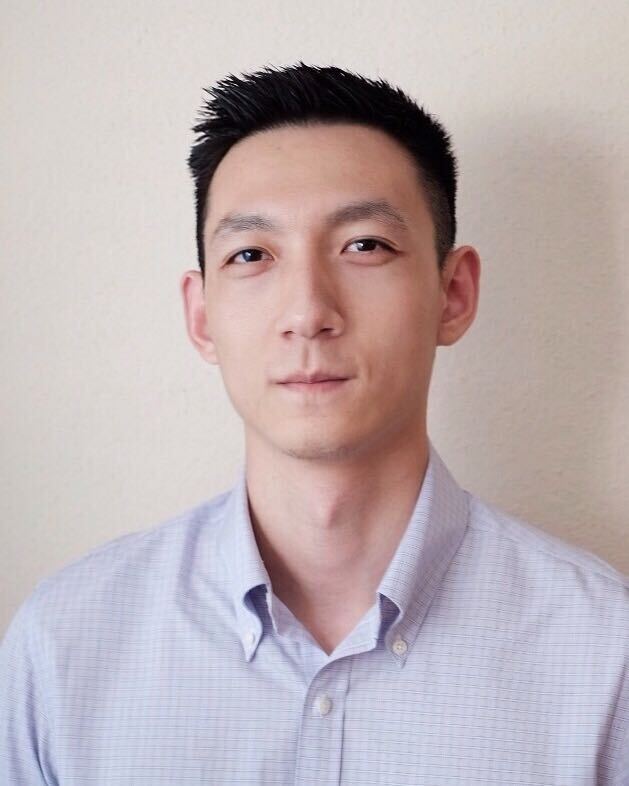}}]{Zhanghao Yu} received the B.E. degree in Integrated Circuit Design and Integrated System from University of Electronic Science and Technology of China, Chengdu, China, in 2016, and the M.S. degree in Electrical Engineering from University of Southern California, Los Angeles, CA, in 2018. He is currently working toward his Ph.D. degree in Electrical and Computer Engineering at Rice University, Houston, TX.

His research interests include analog and mixed-signal integrated circuits design for power management, bio-electronics and security.
\end{IEEEbiography}

\begin{IEEEbiography}[{\includegraphics[width=1in,height=1.25in,clip,keepaspectratio]{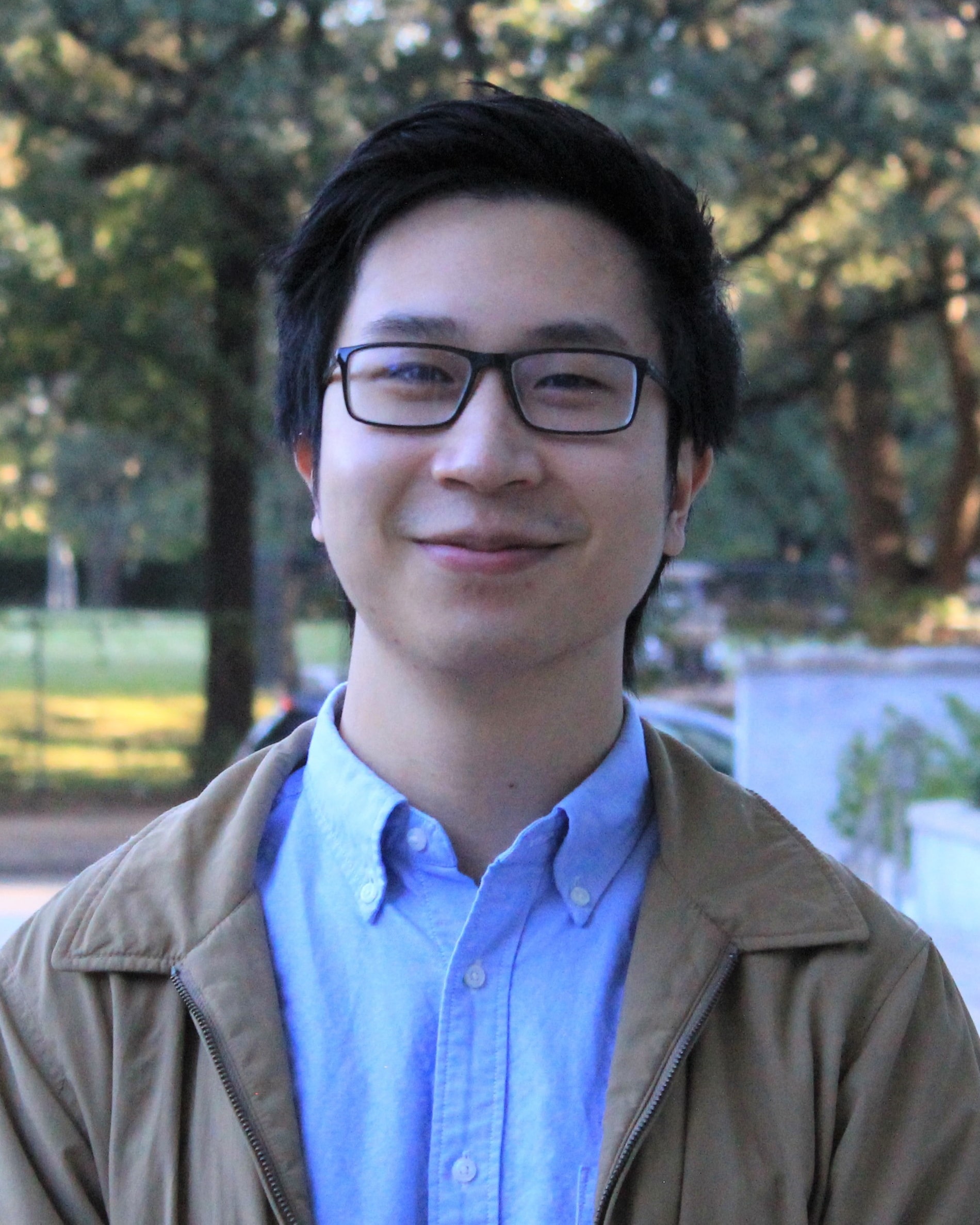}}]{Joshua Chen} received the B.S. degree in Bioengineering from the University of California, Berkeley in 2016. During his time at Berkeley, he was a member of the Berkeley Sensors and Actuators Center (BSAC) where he worked on developing technology for biomedical sensors and 3D printed microfluidics. After graduation, he worked at Verily Life Sciences prior to attending Rice University in Houston, TX to pursue the Ph.D. degree in Bioengineering.

His research interests are in neural engineering, wireless devices, biomedical implants, MEMS devices.
\end{IEEEbiography}

\begin{IEEEbiography}[{\includegraphics[width=1in,height=1.25in,clip,keepaspectratio]{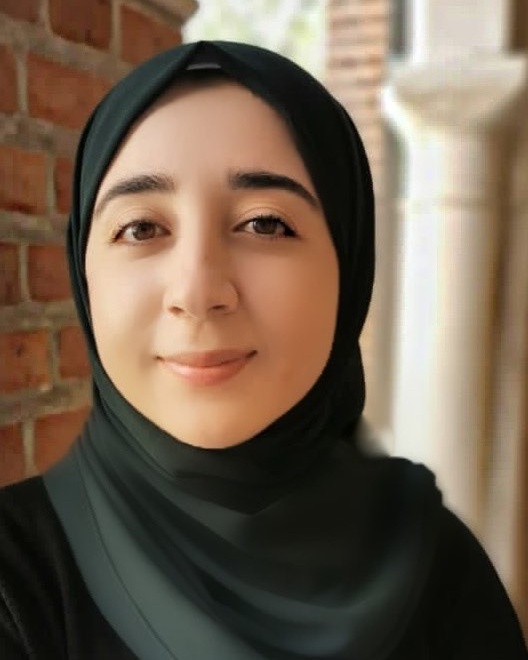}}]{Fatima Alrashdan} received a B.S and M.S degrees in Electrical Engineering from Jordan University of Science and Technology, Irbid, Jordan, in 2016 and 2018, respectively. She is currently working toward the Ph.D. degree in Electrical and Computer Engineering at Rice University, Houston, TX, USA. 

Her research interests include bioelectronics, wireless-implantable biomedical systems, power electronics and control systems for biomedical applications.   
\end{IEEEbiography}

\begin{IEEEbiography}[{\includegraphics[width=1in,height=1.25in,clip,keepaspectratio]{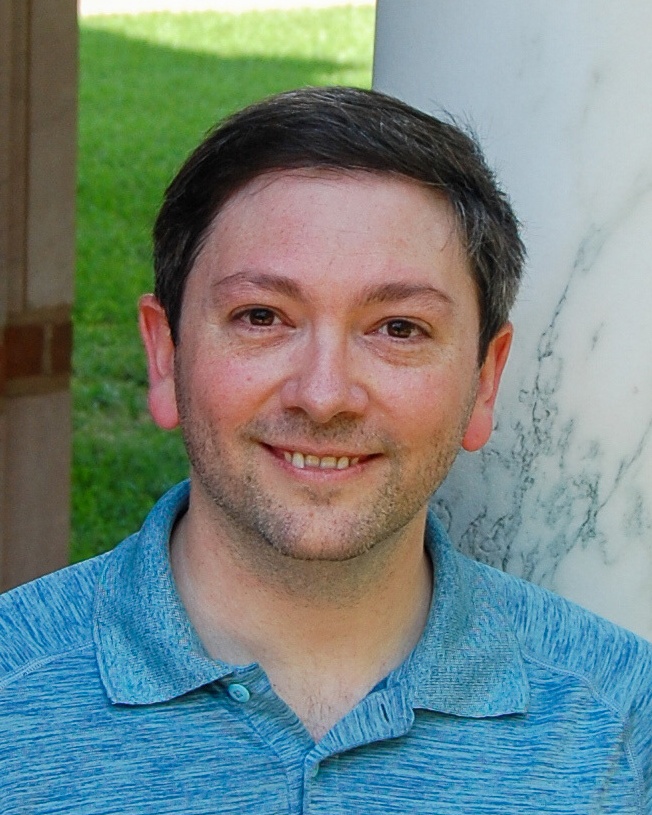}}]{Benjamin Avants} received a B.S. in Electrical Engineering, a B.S. in Computer Engineering, and second majors in Math and Physics from the University of Memphis, Tennessee in 2012. He then joined the Robinson Lab at Rice University, where he has worked as a research engineer ever since.  

Apart from electronics and programming, his interests also include 3D modeling, optics, imaging, and advanced fabrication on the nano, micro, and meso scale.  He enjoys developing researchers' dreams into realities whether it's in the clean room or a maker space.
 
\end{IEEEbiography}

\begin{IEEEbiography}[{\includegraphics[width=1in,height=1.25in,clip,keepaspectratio]{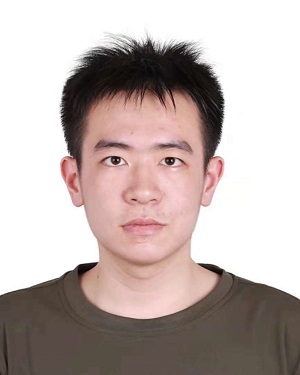}}]{Yan He} received the B.S degree in Electronic Science and Technology from the Zhejiang University, Hangzhou, China, in 2018. He is currently pursuing the Ph.D. degree in Electrical and Computer Engineering with the Rice University, Houston, TX, USA.

His current research interests include analog and mixed-signal integrated circuits design for power management and hardware security.
\end{IEEEbiography}

\begin{IEEEbiography}[{\includegraphics[width=1in,height=1.25in,clip,keepaspectratio]{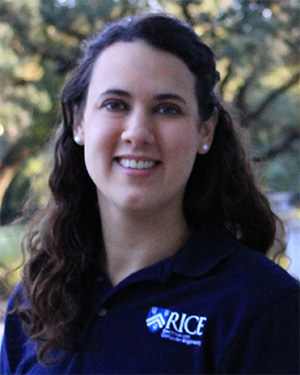}}]{Amanda Singer} received B.S. degree in Physics in 2014 from Saginaw Valley State University, University Center, MI, USA, and a M.A.Sc. degree in Applied Physics from Rice University, Houston, TX, USA, in 2018, and is currently pursuing a Ph.D. degree at Rice University, Houston, TX, USA. 

Her current research interests include the development of wireless bioelectronics and the miniaturization of neural implants.
\end{IEEEbiography}

\begin{IEEEbiography}[{\includegraphics[width=1in,height=1.25in,clip,keepaspectratio]{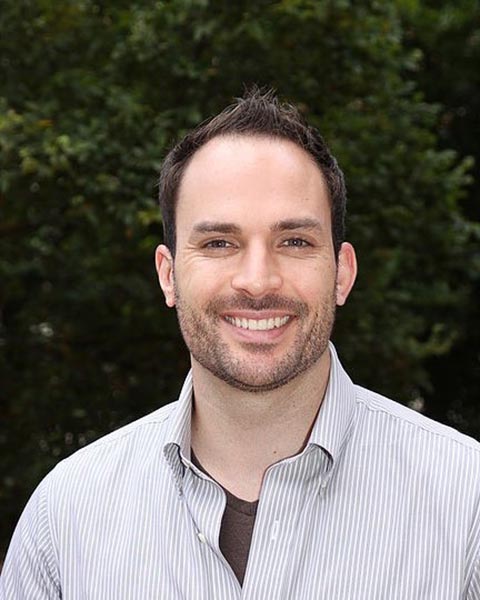}}]{Jacob Robinson}is an Associate Professor in Electrical and Computer Engineering and Bioengineering at Rice University and an Adjunct Assistant Professor in Neuroscience at Baylor College of Medicine. Dr. Robinson earned a B.S. in Physics from UCLA and a Ph. D. in Applied Physics from Cornell. Following his Ph. D., he worked as a postdoctoral fellow in the Chemistry Department at Harvard University. Dr. Robinson joined Rice University in 2012 where he currently works on nanoelectronic, nanophotonic, and nanomagnetic technologies to manipulate and measure brain activity. Dr. Robinson is an IEEE Senior member and currently a co-chair of the IEEE Brain Initiative. He is also the recipient of an NSF NeuroNex Innovation Award, DARPA Young Faculty Award, and Materials Today Rising Star Award.
\end{IEEEbiography}

\begin{IEEEbiography}[{\includegraphics[width=1in,height=1.25in,clip,keepaspectratio]{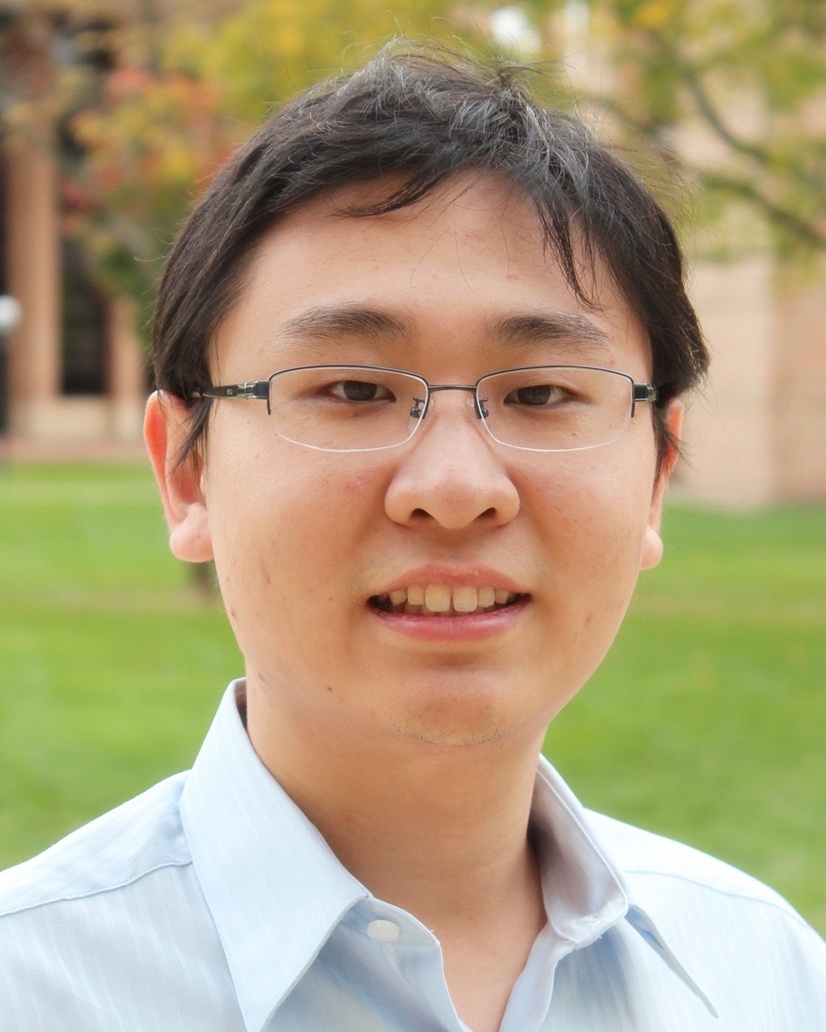}}]{Kaiyuan Yang} (S'13-M'17) received the B.S. degree in Electronic Engineering from Tsinghua University, Beijing, China, in 2012, and the Ph.D. degree in Electrical Engineering from the University of Michigan, Ann Arbor, MI, in 2017. His Ph.D. research was recognized with the 2016-2017 IEEE Solid-State Circuits Society (SSCS) Predoctoral Achievement Award. 

He is an Assistant Professor of Electrical and Computer Engineering at Rice University, Houston, TX. His research interests include digital and mixed-signal circuits for secure and low-power systems, hardware security, and circuit/system design with emerging devices. Dr. Yang received the Distinguished Paper Award at the 2016 IEEE International Symposium on Security and Privacy (Oakland), the Best Student Paper Award (1st place) at the 2015 IEEE International Symposium on Circuits and Systems (ISCAS), the Best Student Paper Award Finalist at the 2019 IEEE Custom Integrated Circuits Conference (CICC), and the 2016 Pwnie Most Innovative Research Award Finalist.
\end{IEEEbiography}



\vfill


\end{document}